\documentclass[10pt, journal, a4paper, final, oneside]{IEEEtran}
\usepackage{amssymb,amsmath,amsthm,epsfig,graphics,subfigure}

\begin{document}

\title{On accurate differential measurements with electrochemical impedance spectroscopy}
\author{S.Kernbach, I.Kuksin, O.Kernbach\\[3mm]
\small Cybertronica Research, Research Center of Advanced Robotics and Environmental Science,\\
\small Melunerstr. 40, 70569 Stuttgart, \emph{serge.kernbach@cybertronica.co}, \emph{igor.kuksin@cybertronica.co}, \emph{olga.kernbach@cybertronica.co}
}
\date{}

\maketitle
\thispagestyle{empty}

\begin{abstract}
This paper describes the impedance spectroscopy adapted for analysis of small electrochemical changes in fluids. To increase accuracy of measurements the differential approach with temperature stabilization of fluid samples and electronics is used. The impedance analysis is performed by the single point DFT, signal correlation, calculation of RMS amplitudes and interference phase shift. For test purposes the samples of liquids and colloids are treated by fully shielded electromagnetic generators and passive cone-shaped structures. Fluidic samples collected from different geological locations are also analysed. In all tested cases we obtained different results for impacted and non-impacted samples, moreover, a degradation of electrochemical stability after treatment is observed. This method is used in laboratory analysis of weak emissions and ensures a high repeatability of results.
\end{abstract}

\section{Introduction}

Electrochemical impedance spectroscopy (EIS) is a common laboratory technique in analytical chemistry \cite{Chang10}, in biological research \cite{Ganesh08}, for example, in the analysis of DNA or structure of tissues, the analysis of surface properties and control of materials \cite{Macdonald06}. This method consists in applying a small AC voltage into a test system and registering a flowing current. Based on the voltage and current ratios, the electrical impedance $Z(f)$ for a harmonic signal of frequency $f$ is calculated. Measured data are fitted to the model of considered system and allow identifying a number of physical and chemical parameters.

\begin{figure}[h]
\centering
\subfigure[]{\includegraphics[width=.35\textwidth]{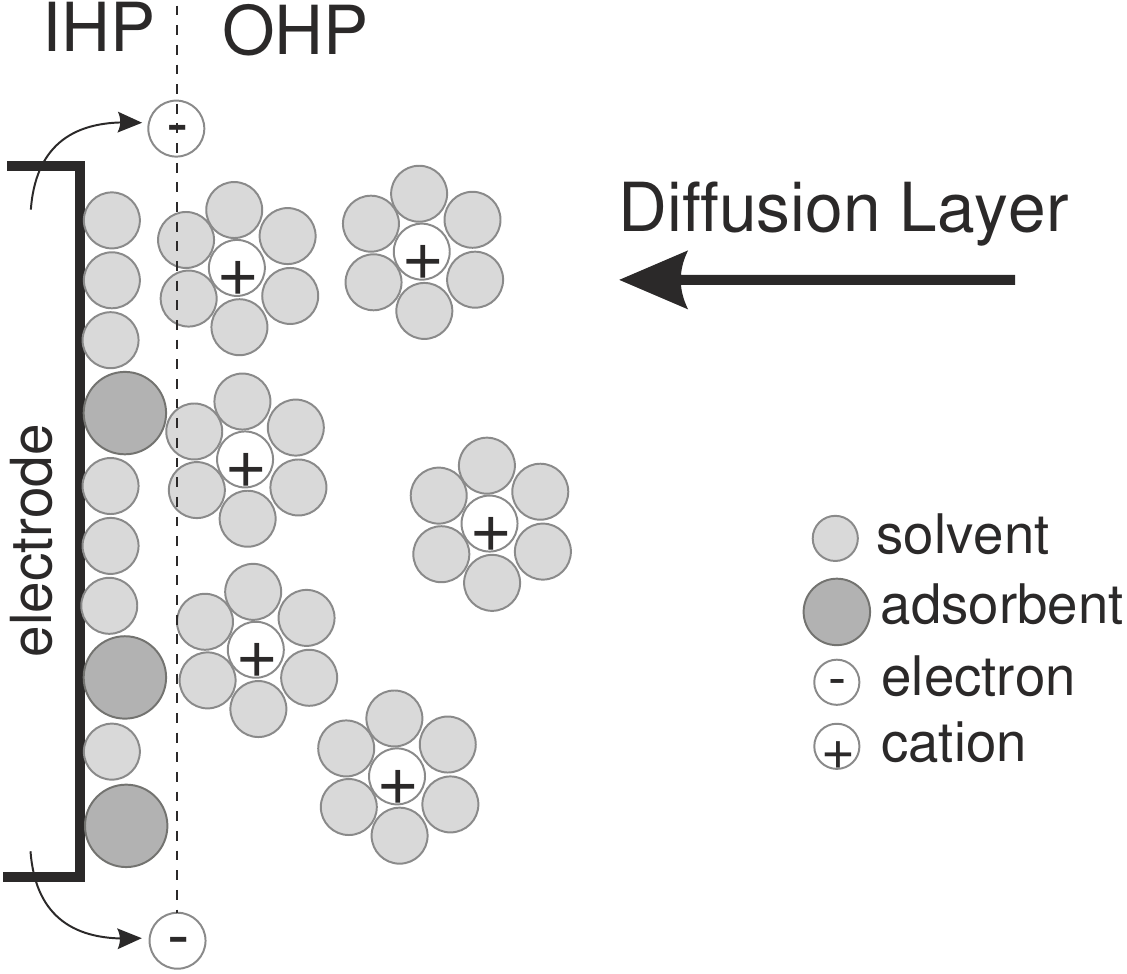}}
\subfigure[]{\includegraphics[width=.35\textwidth]{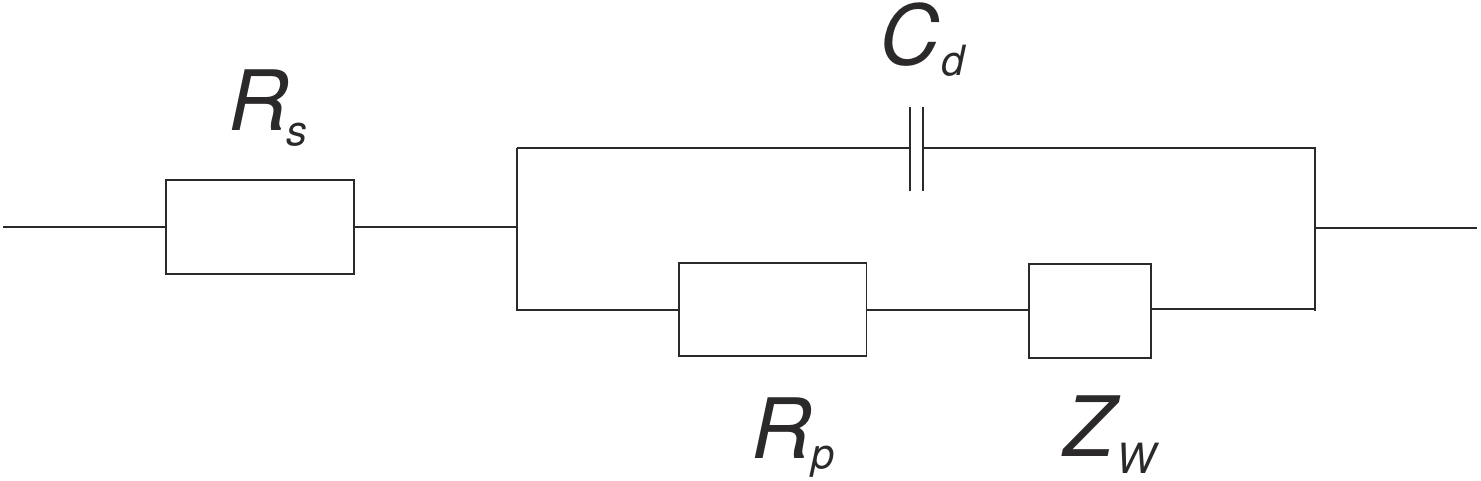}}
\caption{\small \textbf{(a)} Schematic representation of cations, electrons, molecules and adsorbent solutions on the surface of negatively charged electrode, IHP / OHP -- internal and external Hemholtz levels, image from \cite{Chang10}; \textbf{(b)} idealized circuit diagram for the EIS from (a) by following Randles \cite{Randles47}. High-frequency components are shown on the left, low frequency components -- on the right; $ C_d $ -- capacitor on the electric double layer, $R_p$ -- polarization resistance; $R_s$ -- solution resistance, $Z_W$ -- Warburg impedance.
\label{fig:scheme}}
\end{figure}

There are several electrochemical models for EIS. In a number of publications (e.g. \cite{Chang10}) a current flowing through the electrode surface is described by the electrochemical reaction
\begin{equation}\label{eq2}
O + ne^- \rightarrow R,
\end{equation}
where $n$ is a number of transferred electrons, $O$ -- oxidant, $R$ -- reductant. The charge transfer through the electrode surface has Faraday and non-Faraday components. Faraday components appear due to transfer of electrons through the activation barrier and entered into the model as the polarization resistance $R_p$ and solution resistance $R_s$. Non-Faraday current appears due to charge of capacitor on electric double layers close to electrodes. The mass transport of reactants and products causes so-called Warburg impedance $Z_W$ \cite{Chang10}, see Fig.\ref{fig:scheme}.

The state of the art literature describes changes of physico-chemical parameters of solutions, expressed by $R_p$, $R_s$, $C_d$ and $Z_W$, impacted by weak emissions. Sources of such emissions are fully shielded EM (e.g. magnetic vector potential \cite{Puthoff98, AkimovPatent92en}, static electric fields \cite{Burgin08}, LEDs/lasers \cite{Kernbach12JSE}) generators, passive geometrical structures \cite{Kumar05}, \cite{Makin02en}, specific geological locations or other phenomena \cite{Dunne95}, \cite{Schmidt71}, \cite{Tompkins73}. For instance \cite{Bobrov97en}, \cite{Bobrov06en}, \cite{Cardella01} investigated the changes of diffusion Gouy-Chapman layer due to a spatial polarization of water dipoles, see also \cite{Stenschke1985261}, \cite{F29837900225}, \cite{doi:10.1021/la00077a011}. Appropriate electrokinetic phenomena are described by the Gouy-Chapman-Stern model \cite{doi:10.1021/la00077a011}, \cite{Lyklema05}. Papers \cite{Sokolova2002en}, \cite{Andriasheva15en} provided data on the conductivity variation of fluids and plant tissues, measured by the conductometric approach with different frequencies. A number of sources \cite{krasn10en}, \cite{Krinker122en}, \cite{6223212} indicated a change in the of ion transfer of solutions and their detection by potentiometric methods \cite{Kernbach14minimalen}, \cite{Kernbach14dpHen}, \cite{Kernbach15dpHen}. Measurement of various parameters of chemical reactions exposed to weak emissions is well described, for instance, oxidation of a hydroquinone solution and recording the differential absorption spectrum \cite{Anosov03en}, acetic anhydride hydration reaction and recording the optical density
of the solution \cite{Tkachuk10en}, VIS-UV spectroscopy of the acid-base bromothymol indicator and the salt solution $SnCl_2$ \cite{Makin02en}.

Performing multiple measurements of weak emissions by conductometric and potentiometric methods \cite{Kernbach12JSE}, \cite{Kernbach14dpHen}, \cite{Kernbach15dpHen}, we discovered a certain specificity of these measurements. In particular this concerns very small changes of measured values, a high impact of environmental factors, primarily temperature and appearance of phenomena that are not observed in other areas. These works lead to development of new sensitive measuring devices with differential circuits, ultra-low noise, and thermal stabilization of electronics and samples.

In this paper we describe an adapted EIS approach applied to several test systems. Two different EIS-meters are used. The first one is the developed differential EIS-meter with phase-amplitude detection of excitation and response signals, the frequency response is analyzed by a single point DFT and correlation analysis. This system is implemented in hardware in the system-on-chip. The second impedance spectrometer is based on the AD5933 chips from Analog Devices and is used as a control device. It supports only DFT with the Hanning window function. Experiments have shown that changes in samples exposed by weak emissions from fully shielded EM generators, passive geometrical structures and geobiological factors are characterized by four values: the differential signal amplitude (this value is included in all amplitude characteristics obtained by the frequency response analysis); the interference phase shift; ratio between imaginary and real parts of the impedance (as shown e.g. by the Nyquist plot) and a variation of electrochemical stationary of samples. It is assumed that these parameters can indicate changes in near-electrode layers and diffusion processes in the Randles electrochemical model \cite{Randles47}. EIS allows analyzing various liquid and colloidal system and, as an example, we perform analysis of bottled water and milk. Since this work has an explorative character, we do not intend to collect statically significant data for a particular system -- this represents a task for further works.

This paper has the following structure. The section \ref{sec:theory} briefly describes background of EIS,  systematic and random errors and used devices. Sections \ref{sec:results} and \ref{sec:conclusion} consider the obtained results and draw conclusions from these measurements.

\section{Impedance measurement, errors and description of the device}
\label{sec:theory}

There exists an extensive literature on the EIS, both for theory and models, and for technical aspects of measurements. One of the most common methods for measuring impedances is related to an auto-balancing bridge \cite{ImpedanceAgilent}, where a test system is excited by the voltage $V_V$. The flowing current $I$ is converted into a voltage $V_I$ by a transimpedance amplifier (TIA). There are several ways how the signals $V_I$ and $V_V$ are digitized and processed in further analysis.

A common approach consists in analyzing the frequency response (frequency response analysis -- FRA) of the $V_I$ signal, which is based on the discrete Fourier transform (DFT) \cite{Norouzi11} and synthesis of ideal frequencies. This method is sometimes called as the single point DFT \cite{Chabowski15}, \cite{1375091} and requires a fast ADC with 1 msps and more for digitizing the signal $V_I$. The digitized time signal $V_I(k)$ with $N$ samples is converted to a frequency signal $F(f)$, containing real $F_r(f)$ and imaginary $F_i(f)$ parts:
\begin{equation}
\label{eq1}
F_r(f)+iF_i(f)=\frac{1}{N}\sum_{k=0}^{N-1} V_I(k) \left[\cos(\frac{2\pi fk}{N})-i \sin(\frac{2\pi fk}{N})\right].
\end{equation}
It is common to replace $\omega=2\pi f$ and to skip $\frac{1}{N}$, however these parameters are important for calculating the period \cite{Matsiev15}. The magnitude $M(f)$ and phase $P(f)$ are calculated as:
\begin{equation}
\label{eq4}
M(f)=\sqrt{F_r(f)^2+F_i(f)^2}, ~~~P(f)=\tan^{-1}(F_i(f)/F_r(f)).
\end{equation}
Calculation of (\ref{eq4}) is repeated for all $f$ between minimal $f_{min}$ and maximal $f_{max}$ frequencies with the step $\Delta f$. DFT and FRA differ in the way how basic vectors $\cos()$ and $\sin()$ are calculated. In the FRA they are synthesized (for example the sine, see \cite{1375091}):
\begin{equation}
\label{eq3}
u(f)=A\sin(2\pi g(f) f ), ~~~g(f)=\frac{f_{max}-f_{min}}{t_{dur}}+f_{min},
\end{equation}
where $A$ is the maximal amplitude, $t_{dur}$ -- the duration of the measurement, $0<t<t_{dur}$. The synthesized frequency of base vectors must exactly coincide with the frequency of measured signal. In a sense, the expression (\ref{eq1}) calculates a correlation between ideal and measured signals. If $V_V$ and $V_I$ are digitized strictly at the same time points, for example, by using two synchronous ADCs, and since $V_V$ is a sine/cosine signal, (\ref{eq1}) takes the form of correlation between $V_V^f$ and $V_I^f$ (defined for the frequency $f$)
\begin{equation}
\label{eq5}
Corr(f)=\frac{1}{N}\sum_{k=0}^{N-1} V_I^f(k) V_V^f(k).
\end{equation}
Expression (\ref{eq5}) is more preferable because here a non-harmonic signal of high frequency can be used for synthesizing the $V_V$.

The literature discusses non-harmonic signals for driving the electrochemical system, for example, \cite{Ojarand13}, \cite{Mejna09} used digital square waves. An excitation by a broadband noise signal is utilized for a fast impedance spectroscopy \cite{Smith76}. The analysis can be carried out with a direct amplitude and phase detection. For instance, the maximal or RMS amplitude of $V_I$ as well as a phase shift between $V_V$ and $V_I$ can be calculated without FRA. In more complex cases, the Laplace transform is performed between $V_V$ and $V_I$.

If the magnitude or the RMS amplitude of $V_V$ and $V_I$ are known, the impedance $Z$ for the auto-balancing method is defined as
\begin{equation}
\label{eq:cal1}
Z(f)=r_k(f) R_{TIA}\frac{V_V(f)}{V_I(f)},
\end{equation}
where $R_{TIA}$ is a reference resistor in TIA, $r_k(f)$ is a frequency-dependent gain caused by analog circuits, inaccuracies in discrete elements, etc. \cite{ImpedanceAgilent}. Thus, $Z$ is inversely proportional to $V_I$.

\begin{figure}[ht]
\centering
\subfigure[]{\includegraphics[width=.49\textwidth]{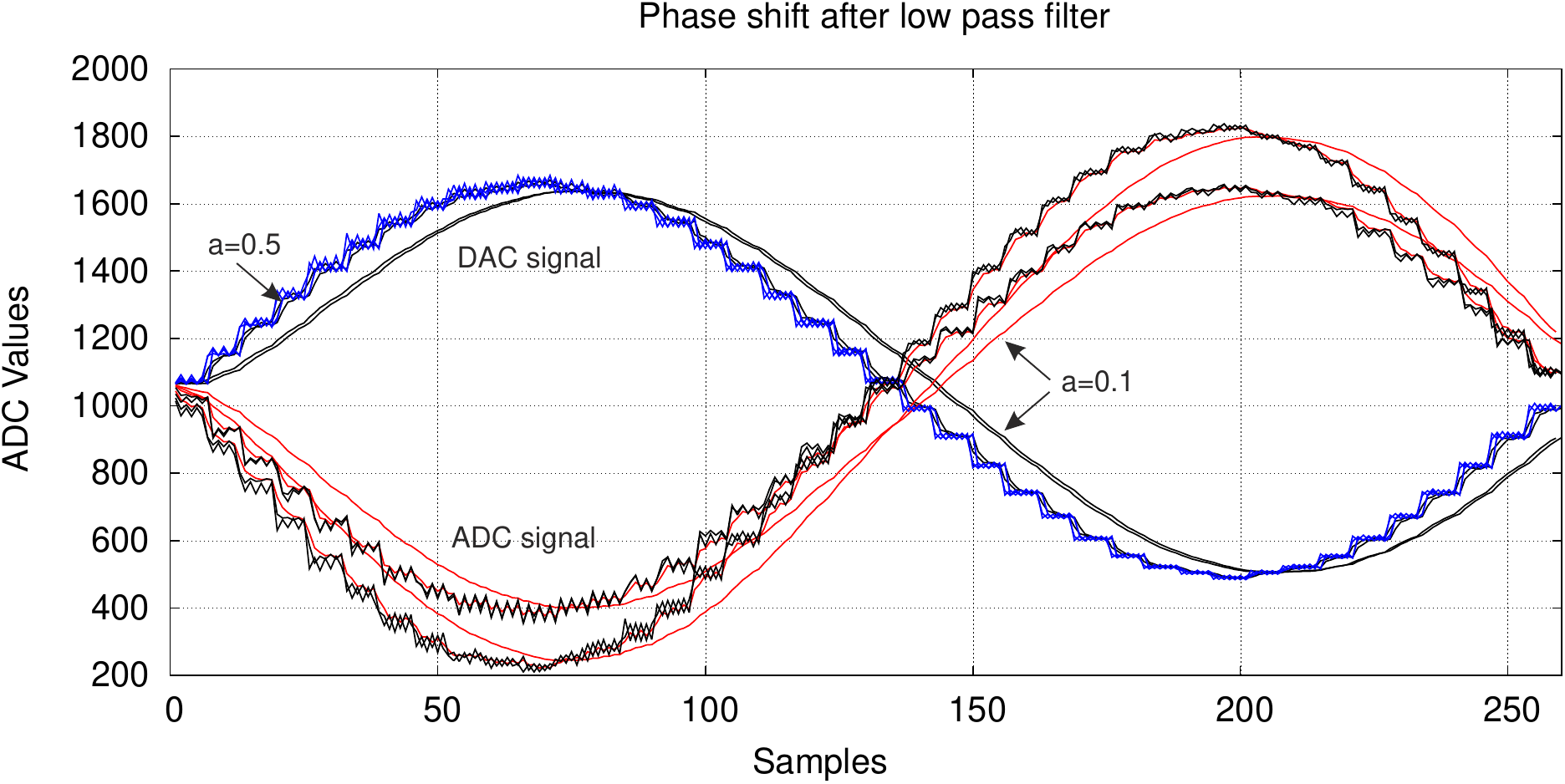}}
\subfigure[]{\includegraphics[width=.49\textwidth]{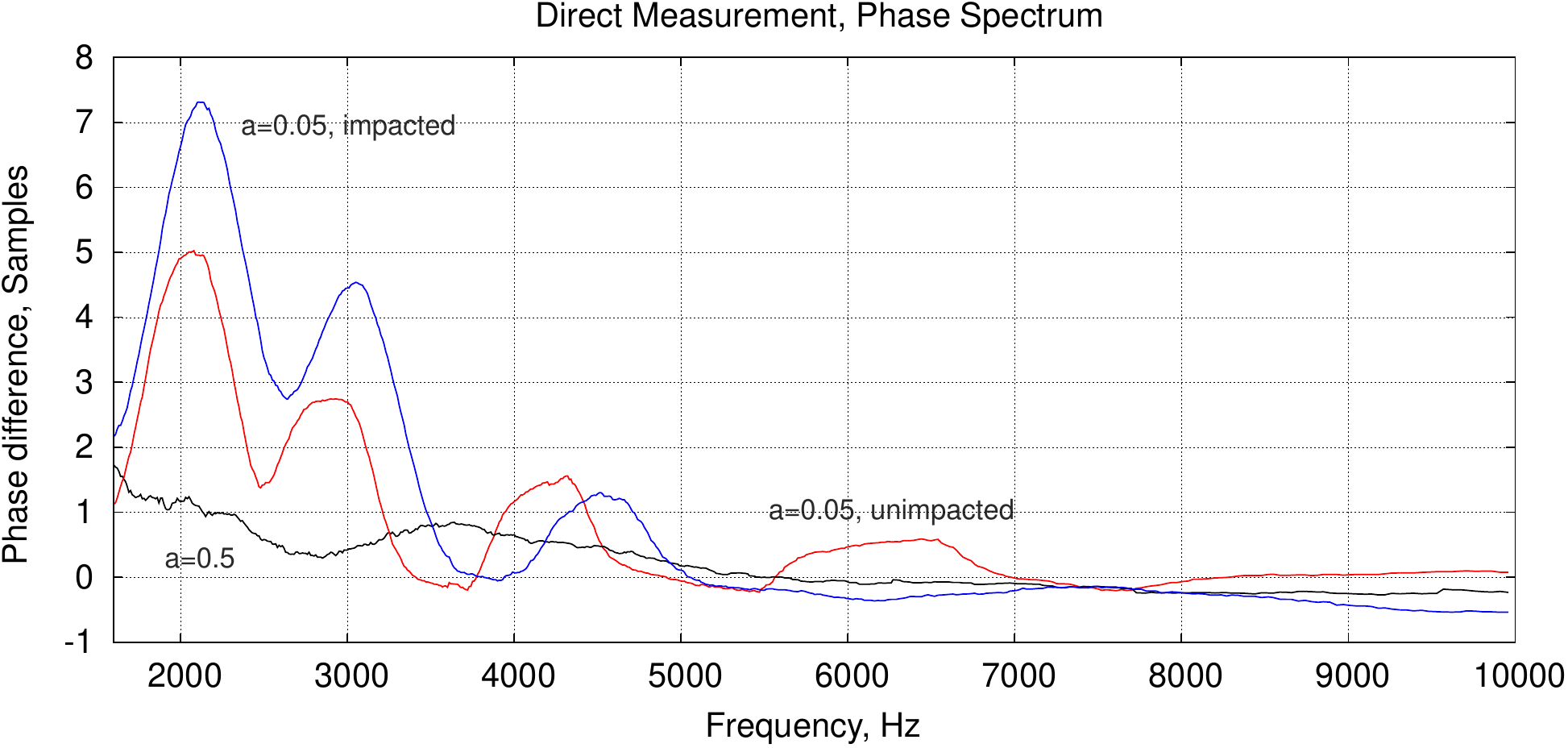}}
\caption{\small The effect of phase variation after applying the digital IIR filter (\ref{eq:digFilt}) with $a = 0.5$, $a = 0.1$ and $a = 0.05$ for a DDS signal synthesis with a small number of samples for \textbf{(a)} one signal period with 40 samples, 3 kHz; \textbf{(b)} the frequency spectrum up to 10 kHz.
\label{fig:IIRfilter}}
\end{figure}

Performing measurements, we found an interesting effect of changing the phase of differential signal $(V_V-V_I)$, as shown in Figs. \ref{fig:calibrationDP}, \ref{fig:measurementWater2DMP}. This unexpected result is not shown by the phase analysis with FRA (\ref{eq4}). This phase variation is created by a combination of two factors: the use of direct digital synthesis (DDS) with a low number of samples and application of the digital infinite impulse response filter (IIR filter):
\begin{equation}
\label{eq:digFilt}
y_{n}=a x_n + (1-a) y_{n-1}=y_{n-1} + a (x_n - y_{n-1}),
\end{equation}
where $x_n$ are discrete samples of an input signal, $y_n$ is the filtered signal, $a$ is a coefficient. Figure \ref{fig:IIRfilter} shows the use of a filter for one period of DDS signal with 40 samples and the frequency spectrum up to 10 kHz.
Such DDS synthesis includes high frequency components (up to 400-600 kHz) in a low frequency signal. Both low- and high-frequency components impact the samples, this is similar to using a broadband noise for a fast spectroscopy \cite{Smith76}. The IIR filter at $a<0.5$ increases the delayed term $y_{n-1}$, however it appears differently at different frequencies. Thus, we assume that the main reason for the periodic phase shift is an interference with the high-frequency components in the DDS signal. Hereinafter, this parameter will be referred to as an interference phase shift $\Phi$. Treatment of samples modifies the response to high frequency components, which is manifested as a variation of $\Phi$. With increasing the frequency, this effect decreases, we observe also a smaller variation of $\Phi(f)$.

As emphasized in \cite{Chang10}, FRA can only be used for stable and reversible systems in dynamic equilibrium, thereby the linearity and stability of electrochemical system should be provided. These are so-called steady-state conditions, whereby deviations from them can appear, for example, in the form of long transient processes. The stationarity is usually not investigated by impedance spectroscopy, but it can represent an additional source of information about the electrochemical processes before and after the treatment of samples.

The measurement approach for weak emissions is described in \cite{Kernbach15dpHen}, \cite{Kernbach13metrologyen}, \cite{Kernbach14dpHen}. It uses two identical channel A (experimental) and B (control), which apply the same exciting signal $V_V$. Spectrograms A and B are subtracted from each other, the resulting difference spectrum is close to zero at all frequencies if the samples A and B are equal. A non-zero differential spectrum indicates differences in samples. Since both samples are prepared at the same temperature, in the same EM, light and other conditions, the difference is caused only by exposure to experimental factors. This method requires at least two measurements. In the first one the samples A and B are measured before impact, these data are used for calibration. The second measurement of A and B is performed after exposing the sample A. The measurement result consists of two differential curves: the calibration (close to zero) and experimental ones. The difference between them allows making conclusions about impact on the sample A.

\subsection{The error analysis}
\label{sec:deviation}

EIS has several systematic and random errors. The first systematic error is related to the period and phase of synthesized base vectors and measured signal. The period of signals accumulated in arrays $V_V$ and $V_I$ must be expressed by an integer. If this condition is not met, so-called leakage errors occur. This problem is solved in three ways \cite{Matsiev15}. Firstly, it is proposed to select the sweep frequencies $f$ in such a way that $V_I$ always contains an integer number of period $k$. The paper \cite{Chabowski15} considered a choice of $f$ based on
\begin{equation}
\label{eq:sampling7}
T_s=k\frac{2^{27}}{f\frac{MCLK}{4}},
\end{equation}
where $T_s$ is the digitization time, MCLK is the fundamental frequency of AD5933 (\cite{Chabowski15} is written in the context of this scheme). The obvious drawback of this approach is that the frequency step $\Delta f$ is large and it is impossible to perform a detailed frequency scan of the test system.

\begin{figure}[ht]
\centering
\subfigure{\includegraphics[width=.49\textwidth]{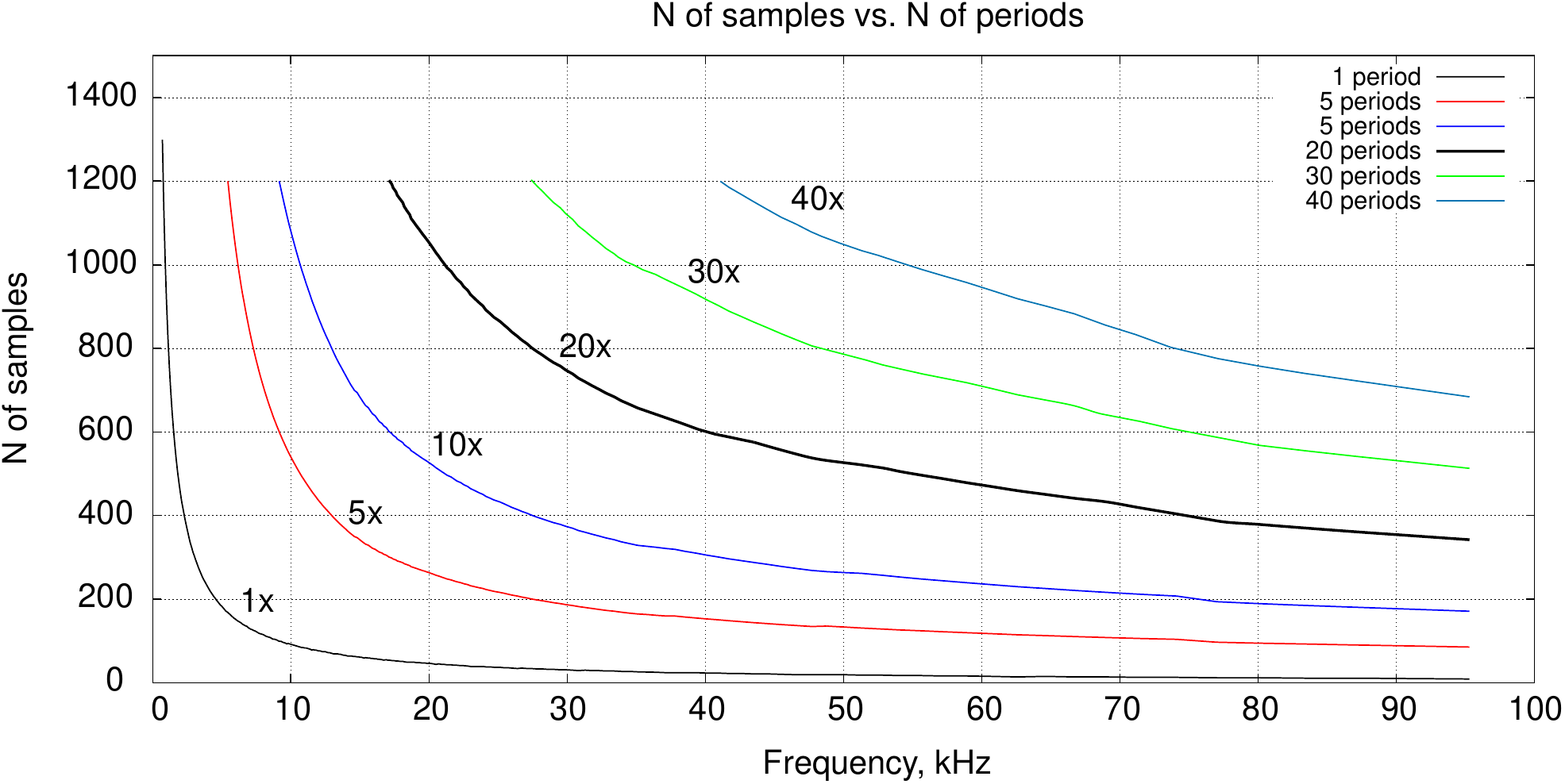}}
\caption{\small The relationship between the frequency $f$ and the number of samples $N_0$ by selecting a different number of periods in $V_I$.
\label{fig:period}}
\end{figure}

The second method is based on adapting the number of samplings $N$ at each sweep. The number of samples $N$ in $V_I$ is changed to $N_0$ in (\ref{eq1}) so that exactly one period of $V_I$ is stored at each reading:
\begin{equation}
\label{eq:sampling}
F(f,N_0)=\frac{1}{N_0}\sum_{k=0}^{N_0}{V_I(k) e^{-i 2 \pi \frac{ f k}{N_0}}}.
\end{equation}
Since the base vectors at FRA must have the same frequency as $V_I$, which is a response to the excitation signal $\sin()$ in $V_V$ (written as $\sin^{V_I}( )$), this leads to
\begin{equation}
\label{eq:sampling2}
F(f,N_0)=\frac{1}{N_0}\sum_{k=0}^{N_0}{\sin^{V_I}(2\pi \frac{f k}{N_0}) e^{-i 2 \pi \frac{ f k}{N_0}}}
\end{equation}
or components-wise, e.g. for the real part of $F_r$
\begin{equation}
\label{eq:sampling3}
F_r(f,N_0)=\frac{1}{N_0}\sum_{k=0}^{N_0}{\sin^{V_I}(2\pi \frac{ f k}{N_0}) \cos(2 \pi \frac{ f k}{N_0})}.
\end{equation}
Since we record only one period of signal, $N_0 = f(f)$ is valid for all frequencies and the main variation of $F_r(f(f))$ occurs due to $N_0$
\begin{equation}
\label{eq:sampling4}
F_r(f(f))=\frac{1}{f(f)}\sum_{k=0}^{f(f)}{\sin^{V_I}(2\pi \frac{k}{N_0}) \cos(2 \pi \frac{k}{N_0})}.
\end{equation}
Despite $f$ is disappeared from $\sin()$ and $\cos()$, $V_I$ and $V_V$ are still affected by the test system at frequencies $f$. Therefore, the physical meaning of (\ref{eq:sampling4}) consists in analysis of amplitude variation, phase and waveform of $V_I$ at the frequency $f$. The disadvantage of this method lies in the rapid decrease of $N_0$ when increasing the frequency $f$. Figure \ref{fig:period} shows the relationship between $f$ and the number of samples $N_0$ by selecting a different number of periods in the $V_I$ array. Thus, it is necessary to introduce the measurement ranges with a different number of periods.

The third approach consists in introducing so-called window function $W(N,k)$
\begin{equation}
\label{eq:sampling5}
F_r(f)=\frac{1}{N}\sum_{k=0}^{N}{V_I(k)W(N,k) \cos(2 \pi \frac{ f k}{N}) },
\end{equation}
which modulates the signal $V_I$ and reduces its amplitude to boundaries of window with $N$ samples. The AD5933 uses the Hanning function \cite{Matsiev15}:
\begin{equation}
\label{eq:sampling6}
W(N,k)=\frac{1}{2}\left(1-\cos(\frac{2 \pi k}{N}) \right).
\end{equation}
The functions (\ref{eq:sampling5}) allows keeping $N$ at the same level for all frequencies, which is useful for digitizing the signal. However, this method distorts the original signal, instead of $V_I(k)$ the signal $V_I(k)W(N,k)$ is analyzed. This problem is discussed in literature \cite{Harris78}. Consequences of using (\ref{eq:sampling5}) is reflected in Fig. \ref{fig:RealImag} as an appearance of new periodic components that are not present in the original signal.

The second source of error is the frequency characteristics, and especially the limited bandwidth of analog components. As a result, the magnitude of $V_I$ decreases with $f$ and the magnitude of the impedance $Z\sim\frac{1}{V_I}$ increases. The problem of increasing impedance exists in all EIS-meters, which solve it in different ways. For example, AD5933 requires two-point or multi-point calibration \cite{AD5933} (see Section \ref{sec:calibration}).

The third source of systematic error represents a small number of samples for DDS synthesis of $V_V$ when measuring the test system with a large capacitive component. This leads to peaks in $V_I$ and significantly increases noise, a detailed examination of this effect for AD5933 is given in \cite{Chabowski15}.

The impedance $Z$ of test system and the reference resistance $R_{TIA} $ in TIA should be similar
\begin{equation}
\label{eq:equal}
Z\sim R_{TIA},
\end{equation}
otherwise the TIA can become saturated and the signal $V_I$ is significantly distorted. Systematic errors are also influenced by the electrode polarization, which is a well-known problem of impedance spectroscopy \cite{0957-0233-24-10-102001}, \cite{Kalvoy11}, especially at low frequencies. This effect is markedly manifested when using small-sized electrode and highly conductive liquids.

Random error also has several components. Firstly, a noise introduces a small random error of detecting the phase-amplitude characteristics of $V_V$ and $V_I$. Secondly, the EIS measurement interacts with samples due to the applied voltage and flowing current. When conducting multiple repeat measurements with the same sample, the measured parameters can 'float' -- that introduces an additional error. A large random error occurs at variation of initial conditions for measurements. This includes small changes in the cell constant, temperature variations of containers and the liquid preparation. These small variations between control and experimental samples are well measurable by the exact differential method and can lead to wrong conclusions about the impacted fluid.

\subsection{The EIS device}
\label{sec:device}

\begin{figure}[ht]
\centering
\subfigure{\includegraphics[width=.49\textwidth]{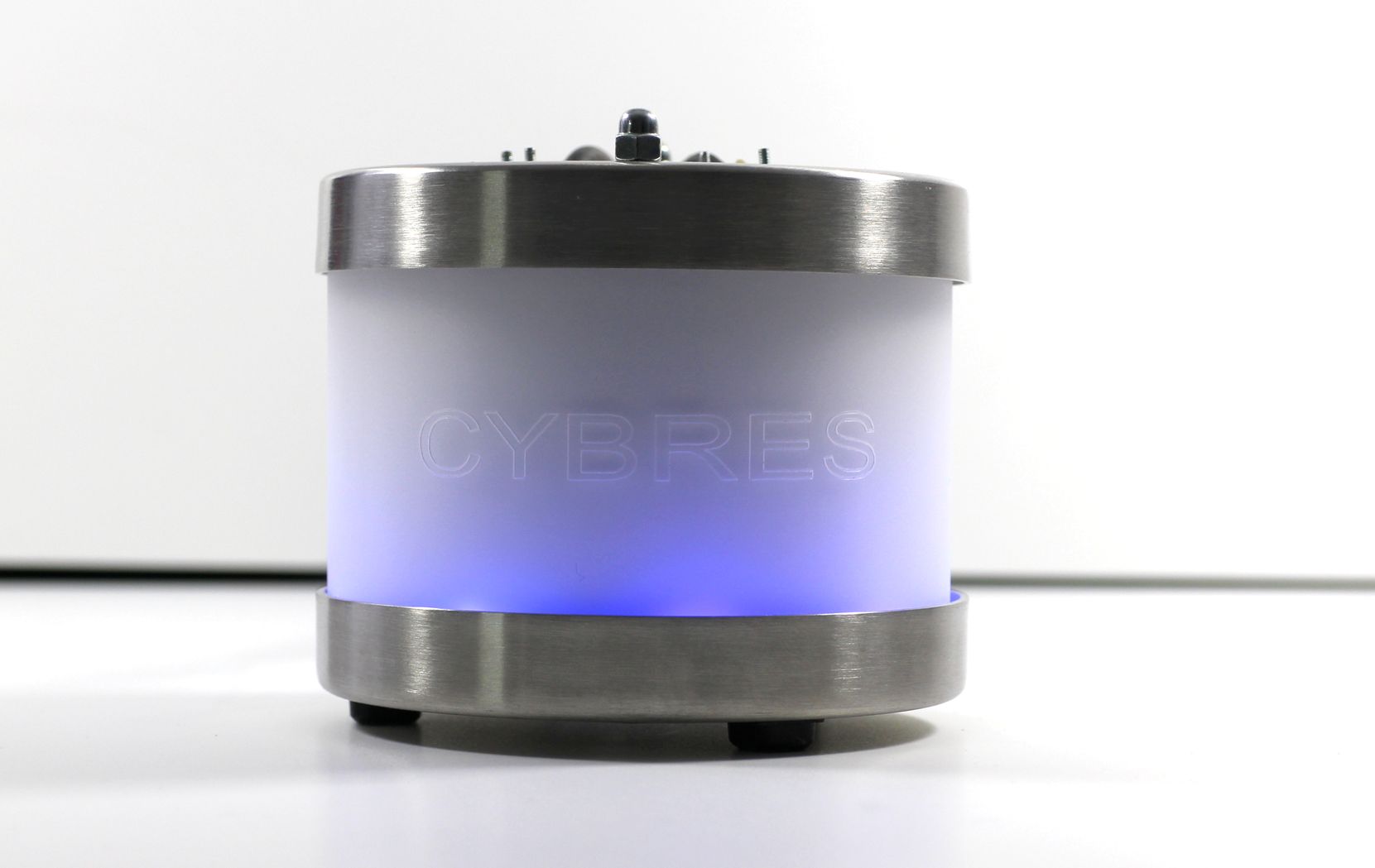}}
\caption{\small Differential impedance spectrometer on MU-EIS system with temperature stabilization of samples and electronic components.
\label{fig:MU}}
\end{figure}

Measurement of weak emissions requires differential measurement circuits and thermal stabilization of system and samples. Available commercial EIS-meters do not offer these options. In the previous works we used commercial conductivity meters, the pH and Redox meters \cite{Kernbach14dpHen}. However they did not provide accurate enough measurements, allowing to characterize weak emissions. In this work we decided to adapt the MU system \cite{Kernbach14dpHen}, \cite{Kernbach15dpHen} for EIS measurements with necessary methodology and metrology, and also to use available devices for control measurements. The first versions of MU-EIS on the PSoC (Programmable System on Chip) architecture were developed in 2012 \cite{Kernbach12JSE}, \cite{Kernbach12ITen} and 2013 \cite{Kernbach13metrologyen}, \cite{Kernbach13formsen} as devices for DC and non-contact high-frequency conductometry. The MU-EIS meter, see Fig. \ref{fig:MU}, supports differential measurements and temperature control, the digital signal processing (DSP core) is implemented in hardware on reconfigurable PSoC architecture.

\begin{figure}[ht]
\centering
\subfigure{\includegraphics[width=.49\textwidth]{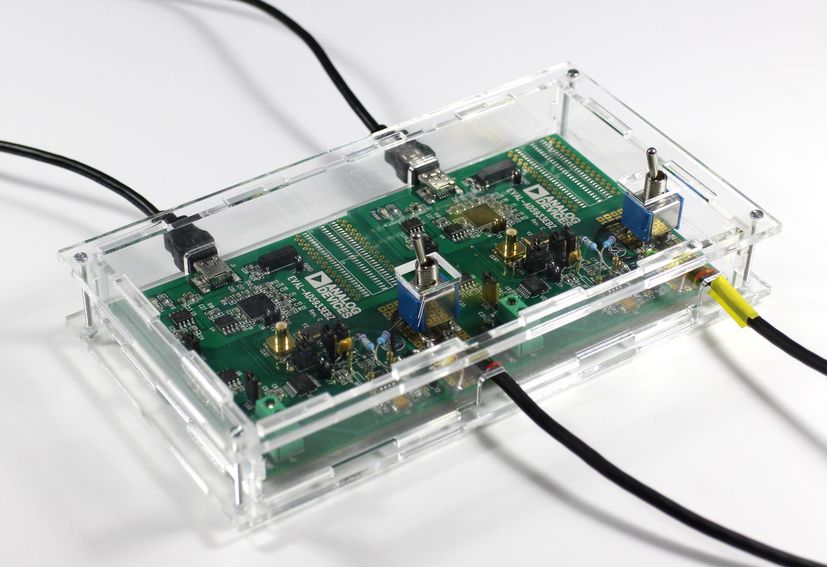}}
\caption{\small Differential impedance spectrometer on AD5933.
\label{fig:AD}}
\end{figure}

The scheme AD5933 \cite{AD5933} has been selected as a commercially available solution. It is a precision impedance spectrometer on a single chip, which has an internal DSP core and is connected to the host system by I2C interface. There are a large number of available devices based on this scheme \cite{Chabowski15}, \cite{Ghaffari15}, \cite{Hoja2010191}. The termostabilization was performed by the MU system, two identical AD5933 boards are used for differential EIS-meter, see Fig. \ref{fig:AD}. Software provided by Analog Devices was rewritten in order to support differential functions.

In general, both EIS-meters are similar. Synthesis of the signal $V_V$ occurs by DDS (AD5933 -- 27-bit frequency resolution, MU-EIS -- 32 bits), the $I-V$ conversion is performed by TIA, the signals are digitalized by 12 bit 1 MSPS SAR ADC (MU-EIS uses two synchronous 1.2 MSPS SAR ADCs for simultaneous sampling of $V_V$ and $V_I$ signals). For impedance matching, both systems use external analog circuitry. There are several fundamental differences between the versions. AD5933 uses the Hanning window function, the number of samples $N$ is fixed on 1024 and the system allows only 512 frequencies $f$ for any measurement range. MU-EIS uses a dynamic adaptation of $N$ within 5 frequency bands, the system allows any number of scanning frequencies. Also, the upper frequency limit in the MU-EIS is 0.6MHz, while the AD5933 is limited by 0.1MHz. MU-EIS allows using non-harmonic signals $V_V$ for driving an electrochemical system, while FRA of the AD5933 does not permit this.

\begin{figure}[ht]
\centering
\subfigure{\includegraphics[width=.4\textwidth]{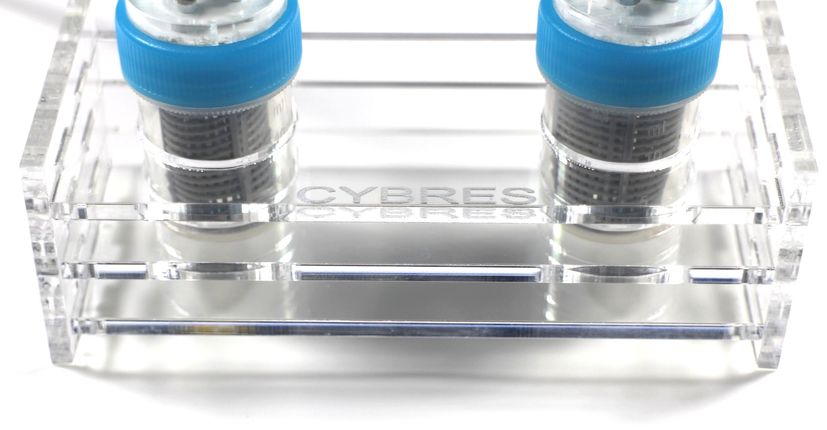}}
\caption{\small Measurement cells with 15 ml containers.
\label{fig:Cells}}
\end{figure}

In both cases the meter is connected to two measurement cells with 15 ml containers. Electrodes (graphite, platinum or stainless steel) are mounted in the upper part of measuring containers, see Fig. \ref{fig:Cells}.

\subsection{Emitting devices}

\begin{figure}[ht]
\centering
\subfigure[\label{fig:cosma}]{\includegraphics[width=.45\textwidth]{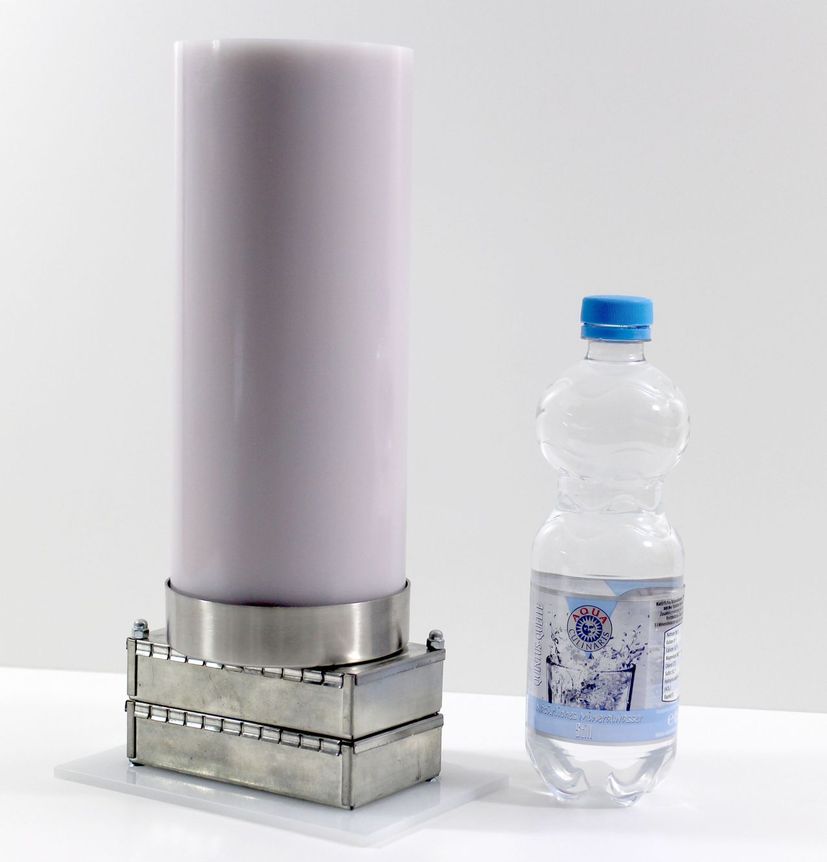}}
\subfigure[\label{fig:contur}]{\includegraphics[width=.49\textwidth]{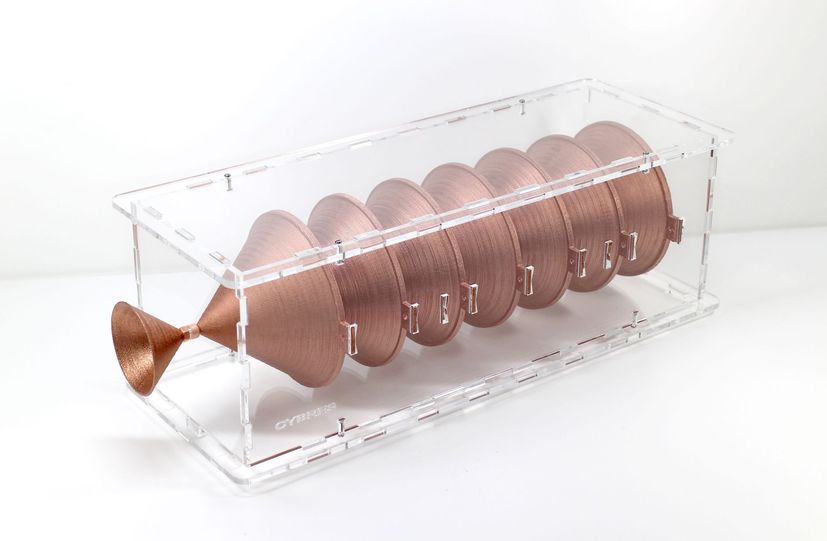}}
\caption{\small \textbf{(a)} Prototype of the \emph{'Cosma'} device used for preparation of liquid samples, suspensions and gels; \textbf{(b)} The system \emph{'Contur'} of passive cone-shaped geometrical structures.
\label{fig:cosmaCon}}
\end{figure}

The device \emph{'Cosma'} shown in Fig. \ref{fig:cosma} was used to prepare the water samples. This device consists of three subsystems that can be switched on or off: LED emitters of ultrashort pulses (based on \cite{Bobrov06en}, \cite{Kernbach12JSE}, \cite{Kernbach12ITen}, the needle emitter of electrostatic field (based on \cite{Veinik81en}, \cite{Veinik91en}, \cite{Chigevsky73en}, \cite{Chigevsky30en}) and the generator of AC magnetic field (based on \cite{Anosov03en}, \cite{Dulnev04en}, \cite{Asheulov00en}, \cite{brit98en}). The generated emission is modulated between 0.1 Hz and 1 kHz. The device also includes a camera for installing a donor substance in order to investigate the imprinting effects \cite{Kernbach15en}. For shielding purposes, emitting elements and electronics are enclosed into grounded metal boxes in the lower part.

A fully passive generator 'Contur' is represented by a system of cone-shaped geometrical forms, see Fig. \ref{fig:contur}. Each cone is made from organic polymer coated from both sides by a copper, each  polymer/copper layer is at least of 0.3mm thick. Cones are placed into each other so that a top of the next cone enters into the previous cone on 1/3 or 1/2 of its height or lies on the baseline. This placement is denoted as the focus position. Experiments are performed with 0\%, 33\% and 50\% focus position and with systems of 3, 4, 5, and 7 cones.

\subsection{Calibration}
\label{sec:calibration}

Calibration of EIS is required to determine the overall gain and to remove nonlinearity and errors introduced by analog components, connecting wires and electrodes. Moreover, a special calibration fluid is required for measuring the cell constant. Attention should be paid to two important points expressed by (\ref{eq:cal1}): firstly, the impedance $Z$ is inversely proportional to $V_I$, secondly, FRA magnitude of $V_V$ is a constant. This allows rewriting (\ref{eq:cal1}) in the form
\begin{equation}
\label{eq:testIm}
Z(f)=r_{total}(f)\frac{1}{V_I(f)},
\end{equation}
where $r_{total}(f)$ is defined by calibration. Another approach consists in calculating the RMS values $V_V^{RMS}$, $V_I^{RMS}$
\begin{equation}
\label{eq:calRMS}
Z(f)=r_k(f) R_{TIA}\frac{V_V^{RMS}(f)}{V_I^{RMS}(f)}.
\end{equation}
Since the value of $R_{TIA} $ is known, the expression (\ref{eq:calRMS}) allows auto-calibration for all measured frequencies $f$ with accuracy of $r_k(f)$. During calibration and measurements it is necessary to set the amplitude of $V_V$ so that the amplitude $V_I$ remains within the operational range of ADC and TIA.

For calibration it is convenient to use the resistor $R$ and the capacitor $C$. The impedance $Z^c$ for a serial connection of $R$ and $C$ is expressed as
\begin{equation}
\label{eq:testIm2}
Z^c=R+\frac{1}{i 2\pi f C},
\end{equation}
where $Z^c_r(f) = R$ and $Z^c_i(f)=\frac{1}{2\pi f C} $. For example, the impedance of the 10nF capacitor at the frequency of 1 kHz is equal to 15915.5Ohm. If the $Z^m (f)$ is the measured impedance at the frequency $f$ for $R$ and $C$, the arrays $k_r(f)$ and $k_i(f)$, represented in the form of
\begin{equation}
\label{eq:testKoef}
k_r(f)=\frac{Z^m_r(f)}{Z^c_r(f)}, ~~~k_i(f)=\frac{Z^m_i(f)}{Z^c_i(f)},
\end{equation}
will contain the calibration coefficients of imaginary and real parts for each frequency. For AD5933  \cite{AD5933} the gain $G$ for calibration impedance $Z^c$ is calculated as
\begin{equation}
\label{eq:eqG}
G=\frac{1}{Z^c M},
\end{equation}
where $M$ is the magnitude (\ref{eq4}). Each measured impedance $Z$ is adjusted by $G$ in the form of
\begin{equation}
\label{eq:eqGM}
Z=\frac{1}{G\sqrt{ (Z_r)^2 + (Z_i)^2}}.
\end{equation}

\begin{figure}[ht]
\centering
\subfigure{\includegraphics[width=.49\textwidth]{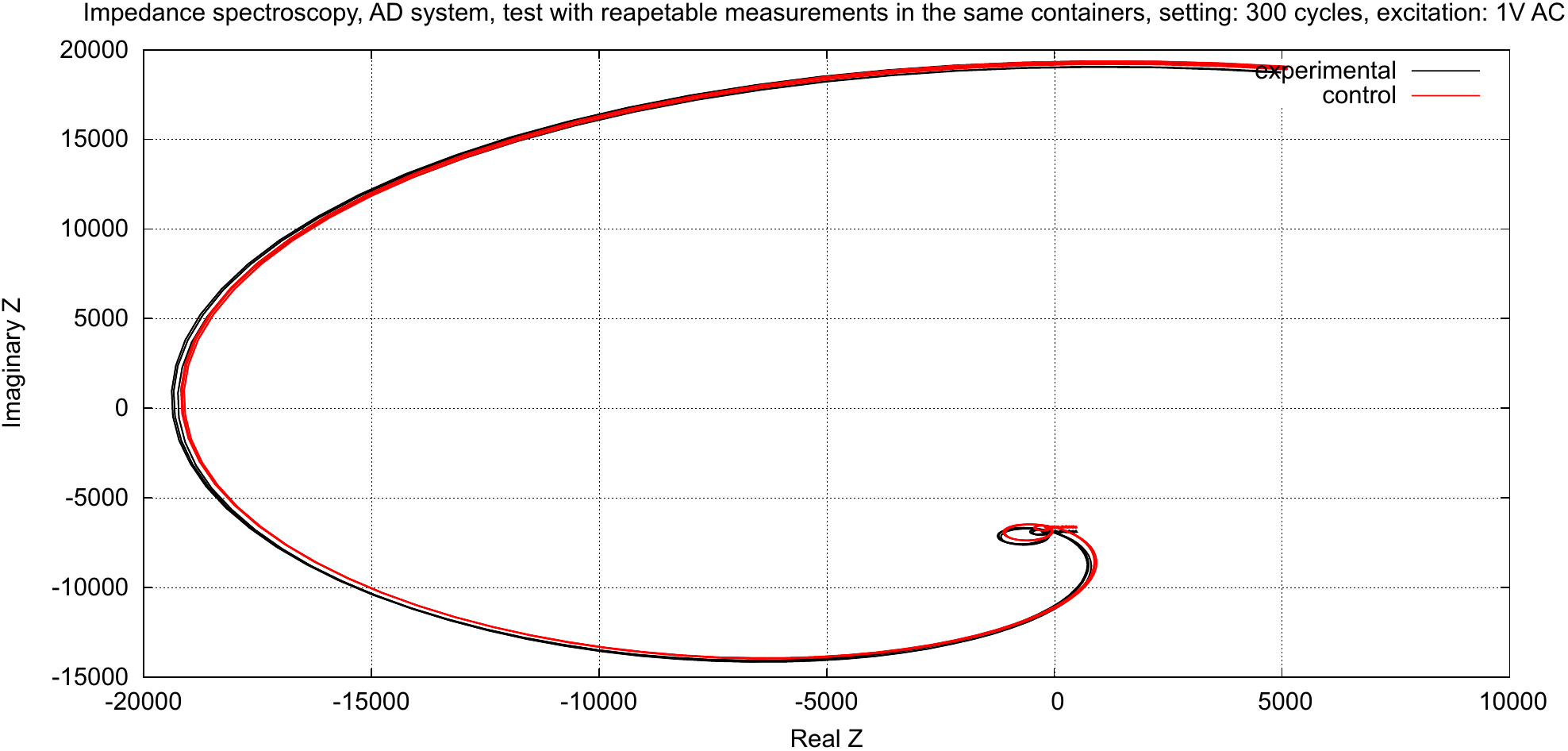}}
\caption{\small The Nyquist plot $Re(Z)$ of $Im(Z)$ for AD5933, oscillating components due to Hanning window function are clearly visible.
\label{fig:calibrationHanning}}
\end{figure}

\begin{figure*}[ht]
\centering
\subfigure[]{\includegraphics[width=.49\textwidth]{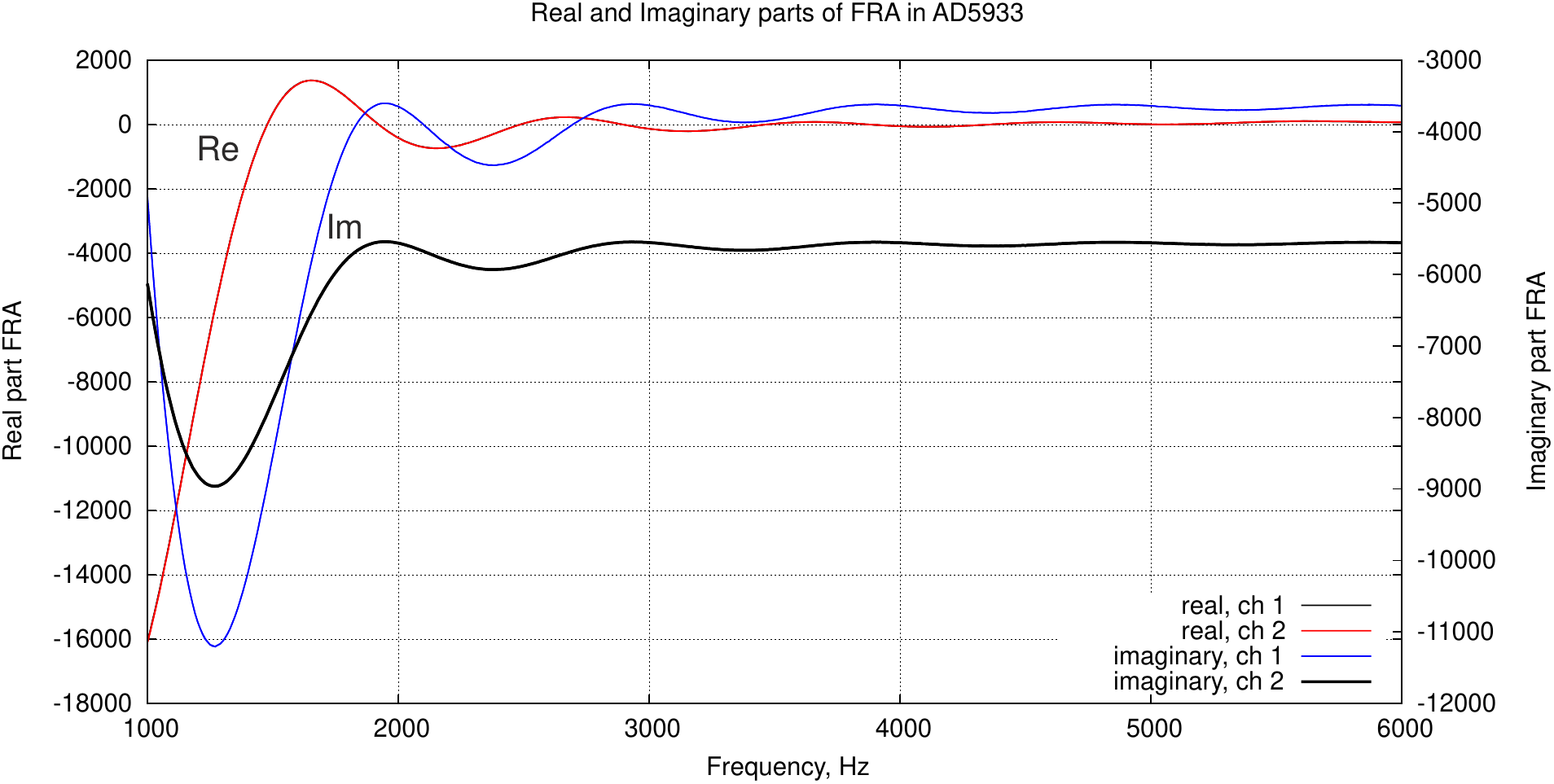}}~
\subfigure[]{\includegraphics[width=.49\textwidth]{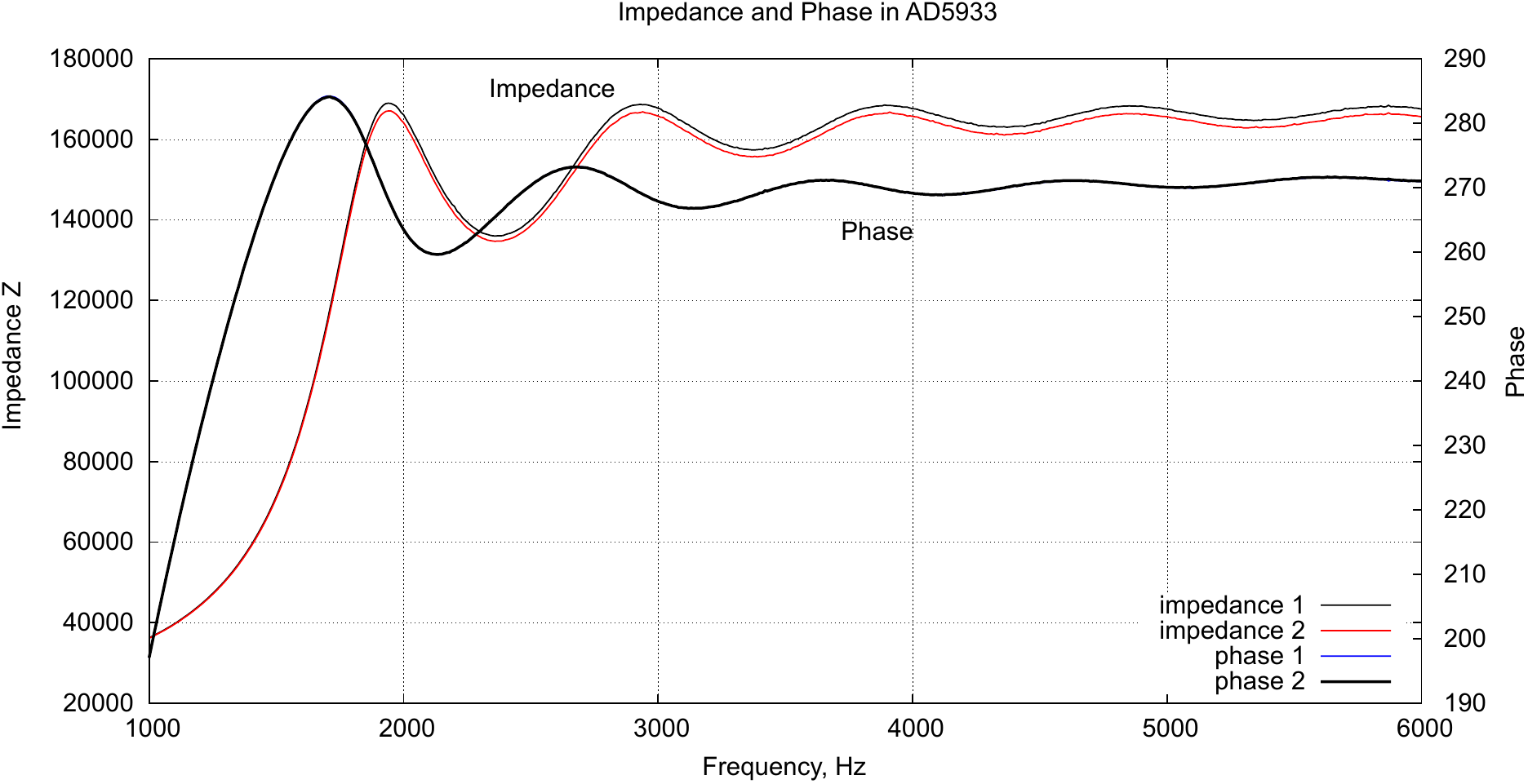}}
\subfigure[\label{fig:RealImagMU}]{\includegraphics[width=.49\textwidth]{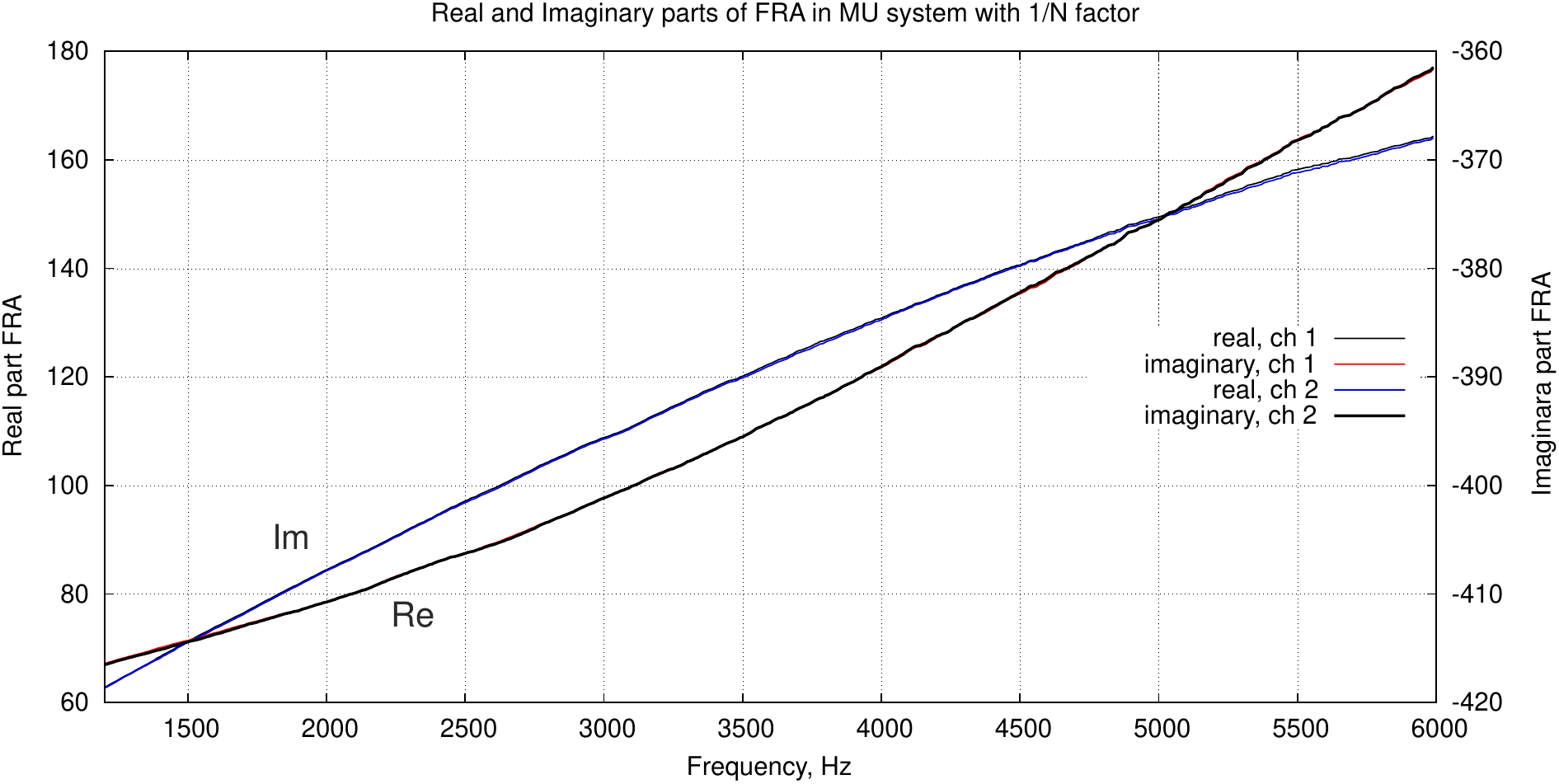}}~
\subfigure[\label{fig:RealImagMUN}]{\includegraphics[width=.49\textwidth]{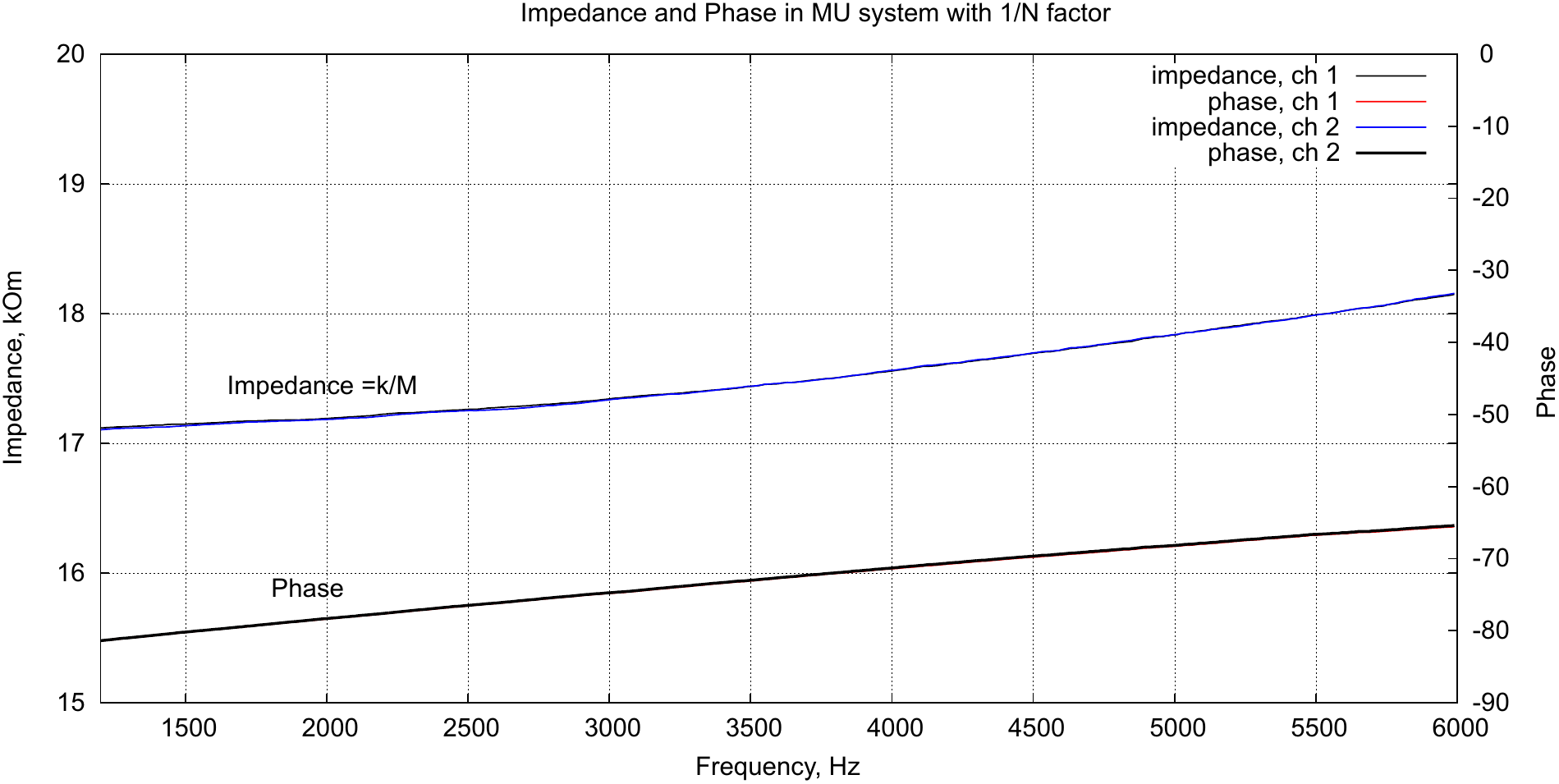}}
\caption{\small \textbf{(a)} Real and imaginary parts and \textbf{(b)} impedance and phase of the FRA obtained for the AD5933; \textbf{(c)} real and imaginary parts and \textbf{(d)} impedance and phase of the FRA obtained with MU-EIS. The measurements were performed with a resistance of 17.6 ohms in steps of 10 Hz with single-point calibration, excitation is performed by a sinusoidal signal.
\label{fig:RealImag}}
\end{figure*}

Figure \ref{fig:RealImag} shows the real and imaginary parts of FRA, as well as the magnitude and phase of the impedance with the calibration resistance of 17.6 kOm. As mentioned above, AD5933 circuit utilizes a window function, the oscillating components of $\sin()$ and $\cos()$ are well visible in the output arrays $Re(Z)$ and $Im(Z)$, see the Nyquist plot in Fig. \ref{fig:calibrationHanning}. To compensate this effect, AD5933 requires a calibration across all frequencies, indicating RC models of the calibrated device.

\begin{figure}[hpt]
\centering
\subfigure[]{\includegraphics[width=.49\textwidth]{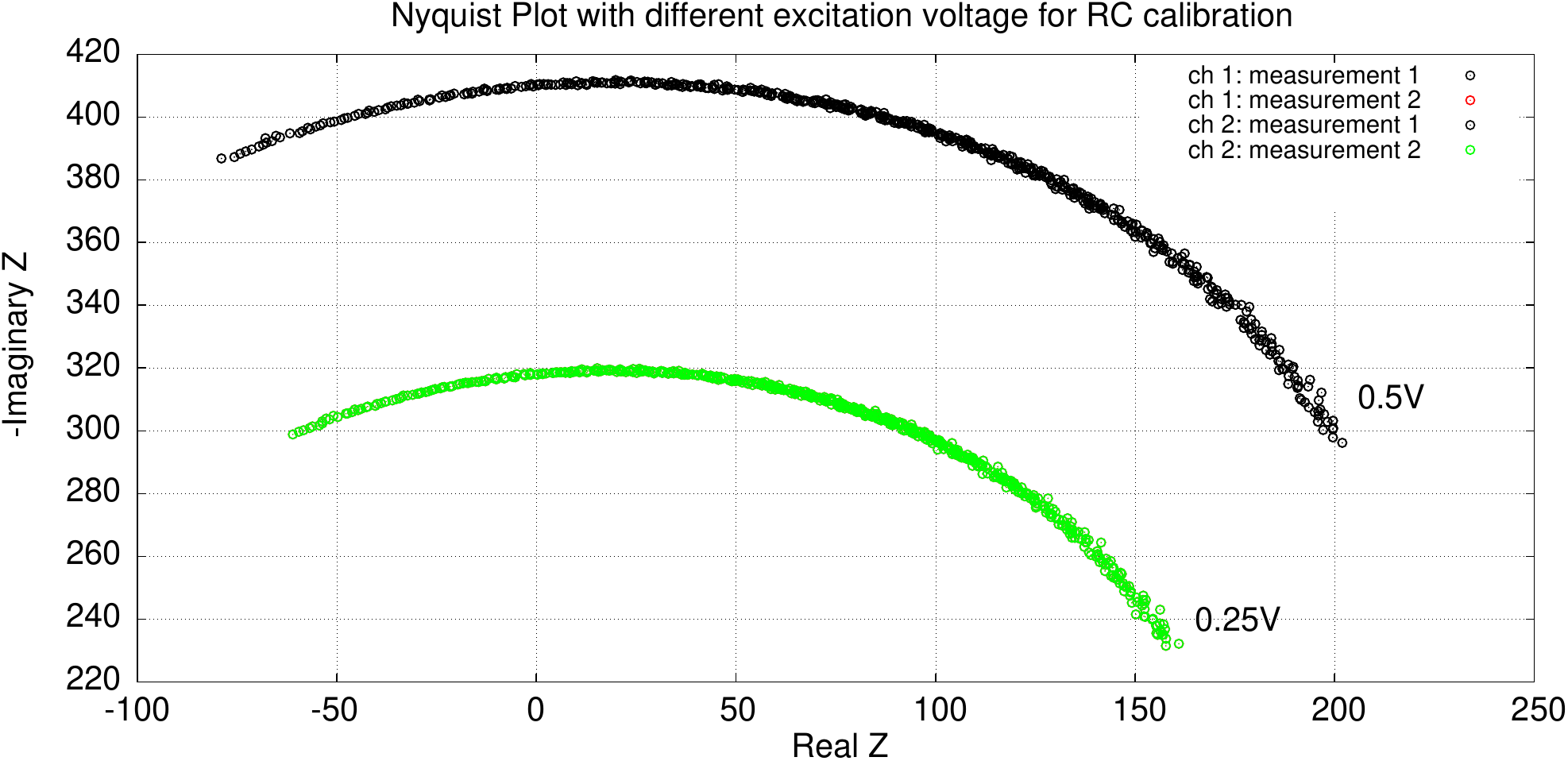}}
\subfigure[]{\includegraphics[width=.49\textwidth]{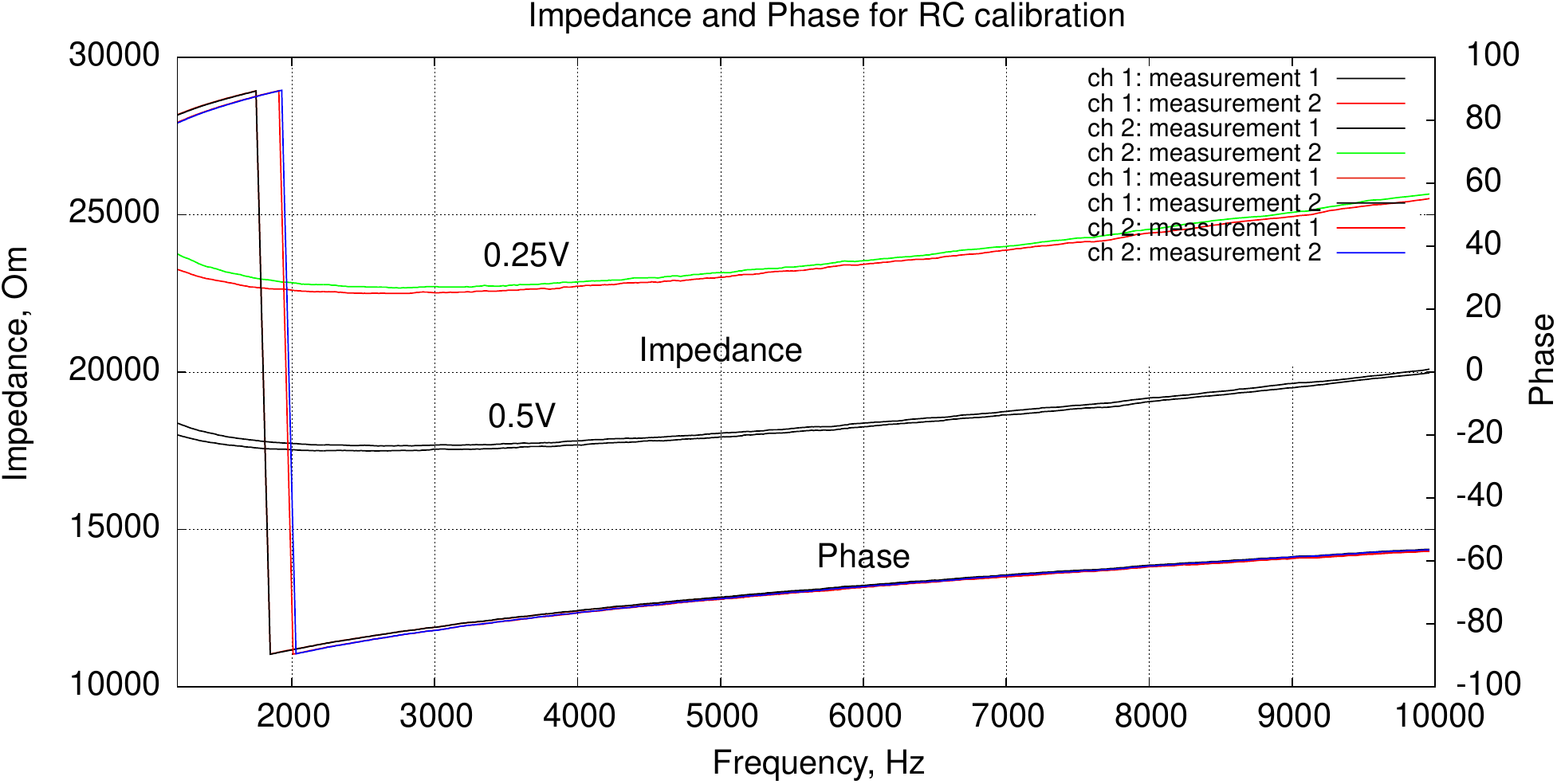}}
\caption{\small \textbf{(a)} Nyquist plot and \textbf{(b)} impedance and phase for the calibration RC-chain (17.6 kOm, 10 nF) with a one-point calibration. Phase jump is due to the function $\tan^{-1}()$ in (\ref{eq4}).
\label{fig:calibrationRC}}
\end{figure}

Figures \ref{fig:RealImagMU}, \ref{fig:RealImagMUN} show results the real and imaginary parts of FRA for MU-EIS as well as the magnitude and phase without oscillations. Phase is about $-90^\circ$ because of TIA converter. The correlation curve is obtained by (\ref{eq5}), it follows the magnitude. Figure \ref{fig:calibrationRC} shows the calibration data for the single-point calibration with $R=17.6$ kOm, $C=10$nF and the obtained Nyquist plot for $Re(Z)$ and $Im(Z)$.

\begin{figure}[htp]
\centering
\subfigure[]{\includegraphics[width=.49\textwidth]{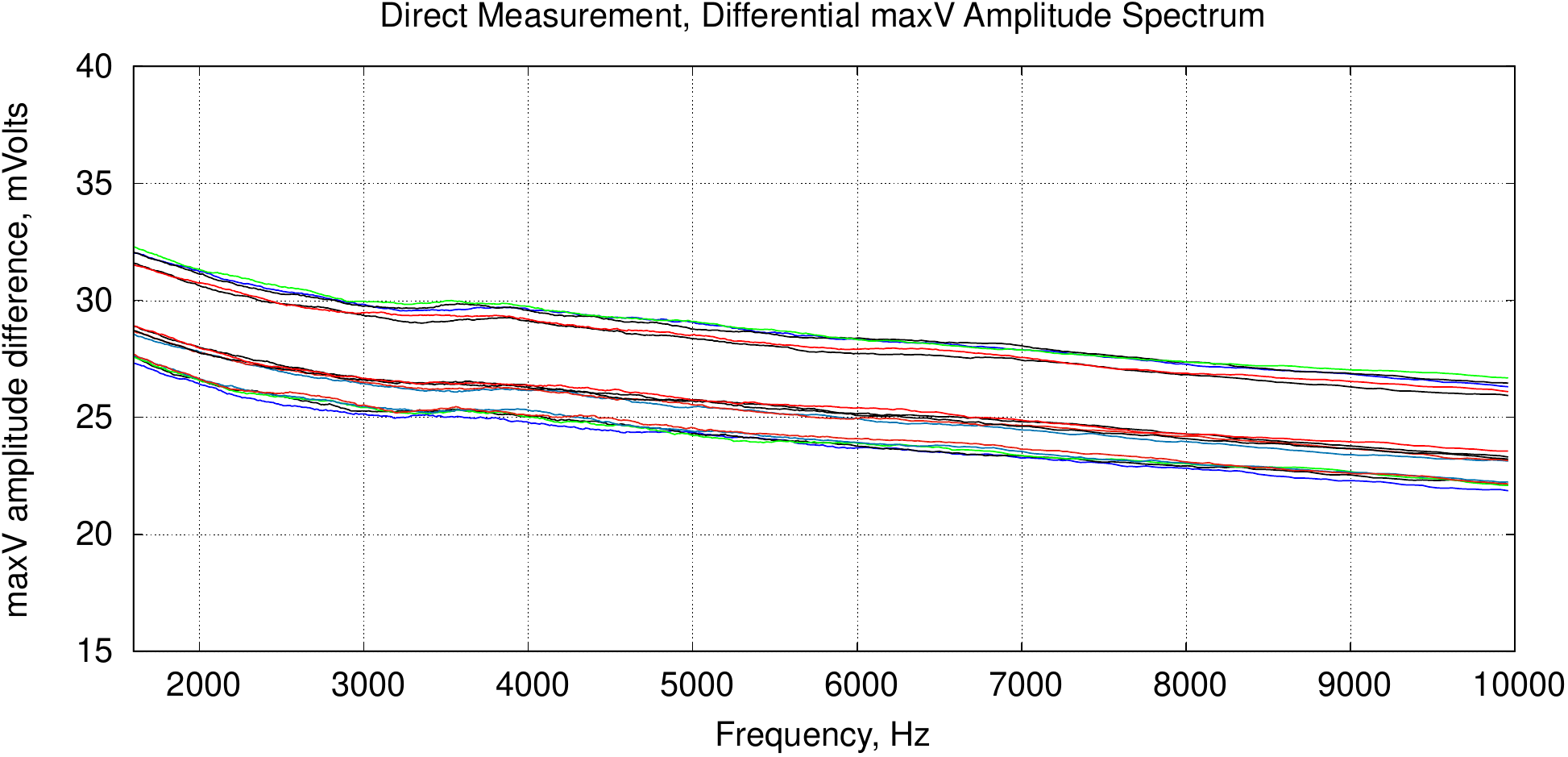}}
\subfigure[\label{fig:calibrationDP}]{\includegraphics[width=.49\textwidth]{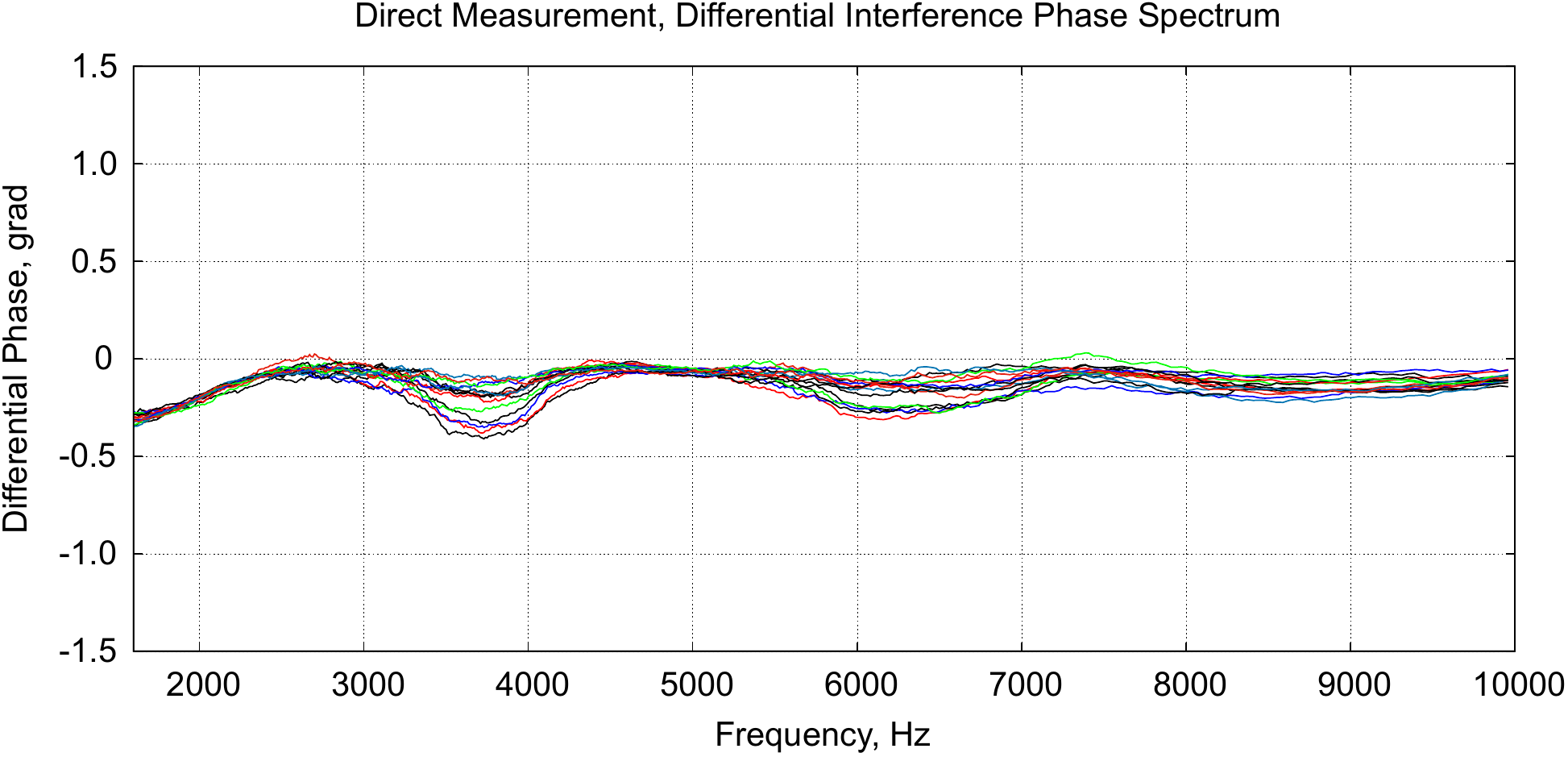}}
\caption{\small Control measurements of the bottled water 'Black Forest', see description in text. \textbf{(a)} The differential amplitude spectrum $V_I$, obtained by MU-EIS; \textbf{(b)} differential phase spectrum by the MU-EIS. 
\label{fig:calibration}}
\end{figure}

\begin{figure}[htp]
\centering
\subfigure{\includegraphics[width=.49\textwidth]{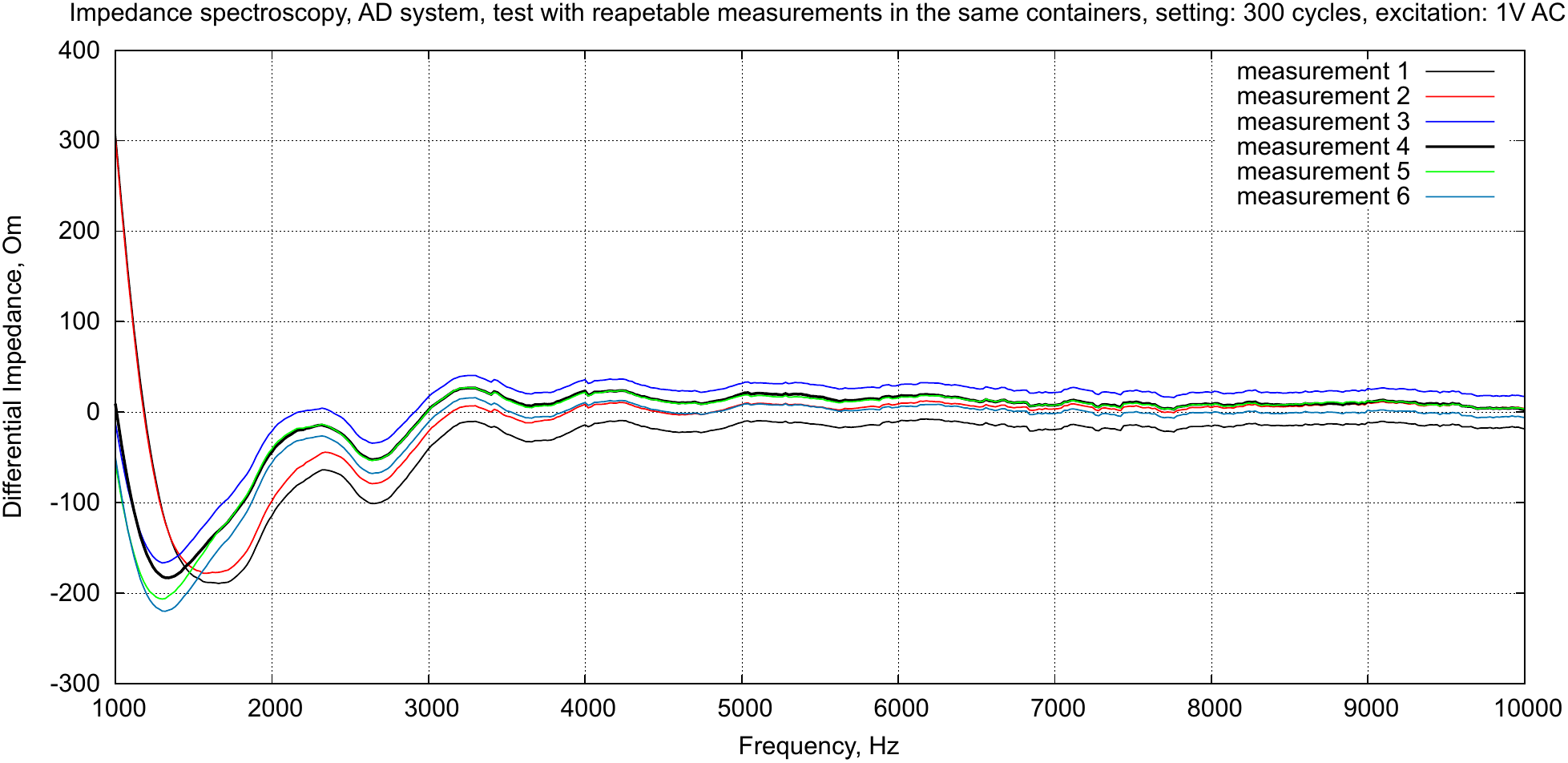}}
\caption{\small Control measurements of the bottled water 'Black Forest' with AD5933, differential impedance spectrum, see description in text. 
\label{fig:calibrationAD}}
\end{figure}

The paper \cite{Kernbach15dpHen} expressed arguments against the calibration of differential measurements because the expressions (\ref{eq:testKoef}) and (\ref{eq:eqG}) introduce additional noise from test measurements. Since the amplitude of differential signal is small, this noise complicates a detection of small changes. For this reason, Section \ref{sec:results} shows the results without calibration.

The cell constant includes factors associated with the geometry of measuring cell and electrodes, the surface area, fluid dynamics in the cell, etc. To analyze the errors caused by variation of the cell constant, the following method is used \cite{Kalvoy11}. First, each measurement starts from dry electrodes and is repeated to determine the reproducibility of measurements. Further, the electrodes are removed and dried. This measurement cycle is repeated again to determine the degree of variability. Figures \ref{fig:calibration} and \ref{fig:calibrationAD} show an example of such measurements with 10 iterations both for MU-EIS, and for AD5933. Results for the same initial conditions do not vary more than $\pm0.1$ ohms (0.001\% of the total value). The variation of initial conditions is between $\pm 5$ Ohm to $\pm20$ Ohm (from $\pm$ 0.05\% to $\pm $ 0.2\% of full scale).

\section{Measurement results}
\label{sec:results}

The experiments are performed in the following way. The bottled water 'Black Forest' at room temperature is poured into four 15 ml containers (two samples A and B) for AD5933 and MU-EIS system. Samples thermostat is set on $27^\circ$C. The containers are kept in thermostat for 20 minutes to equalize the temperature before starting measurements.

\textbf{1.} First test measurements are performed with A and B samples at frequencies between 1 kHz and 10 kHz with 10 Hz steps. Sweeps are repeated 10 times, the spectra of A and B are subtracted from each other, as a result 10 differential spectral curves are obtained. The purpose of this test is to assess the bias error of repeating measurements.

\textbf{2.} To determine the random error, the samples are removed from the thermostat and put on a shelf. After 30 minutes the samples are placed again in the thermostat and 10 differential measurements as described in (1) are performed.

\textbf{3.} The samples A and B are removed from the thermostat, the sample A is treated by \emph{'Cosma'} or \emph{'Contur'} devices. After this, the sample A is rested about 10 minutes, then measurements with both samples are performed again.

\textbf{4.} To determine the random error after exposure, the samples (after exposure of the sample A) are removed from the thermostat, rested for 30 minutes, and then 10 differential measurements are carried out.

Thus, this approach allows evaluating the variations of systematic and random errors as well as to estimate  the effect of exposing the water to experimental factors. In total, 5 series of experiments with multiple iterations are performed.

\begin{figure}[hb]
\centering
\subfigure{\includegraphics[width=.35\textwidth]{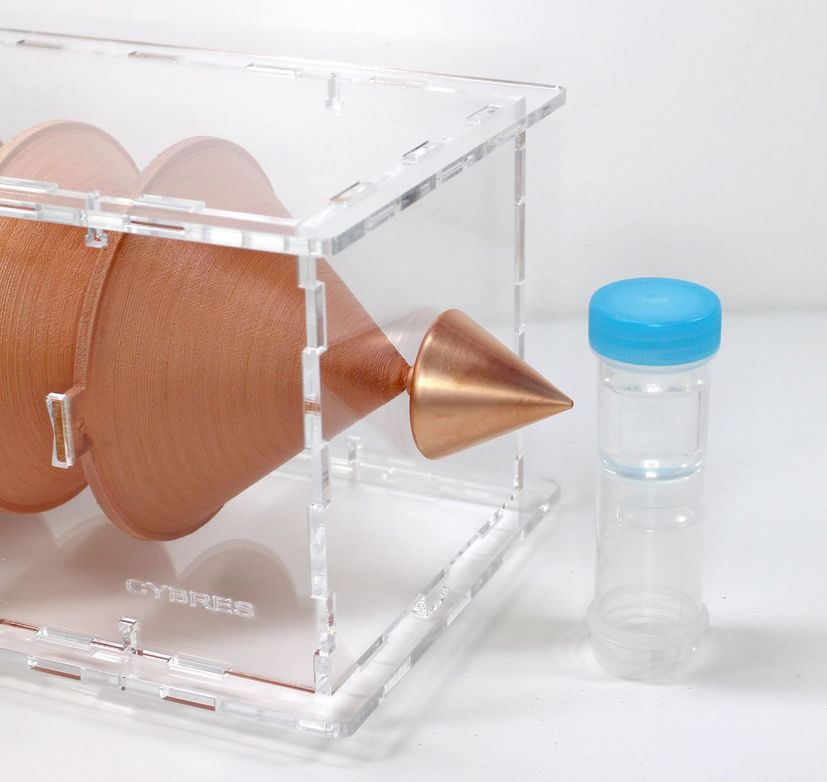}}
\caption{\small Setup with cone-shaped geometric structures and water samples.
\label{fig:coneSample}}
\end{figure}

\begin{figure*}[htp]
\centering
\subfigure[]{\includegraphics[width=.49\textwidth]{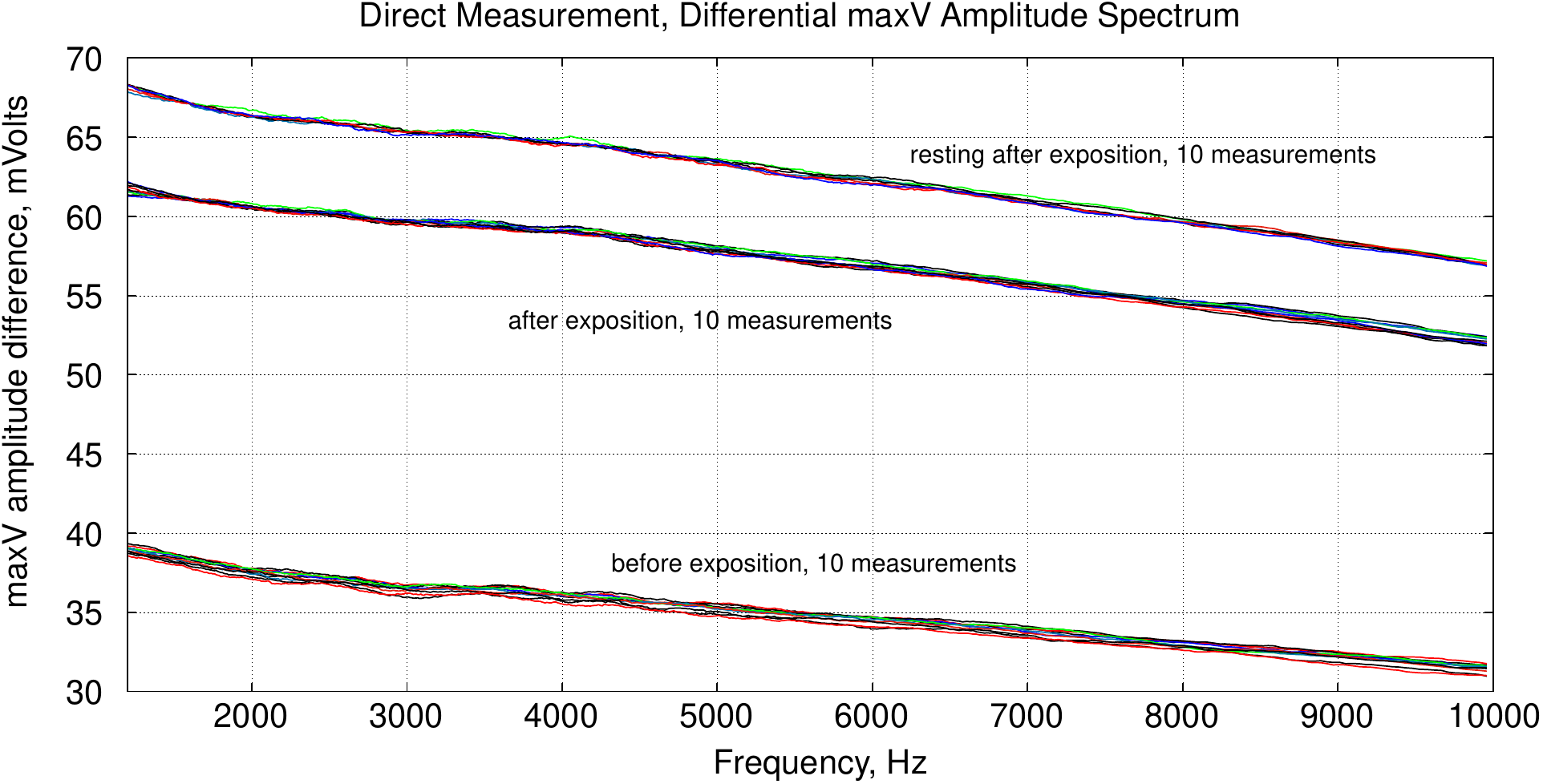}}~
\subfigure[\label{fig:measurementWater2DMP}]{\includegraphics[width=.49\textwidth]{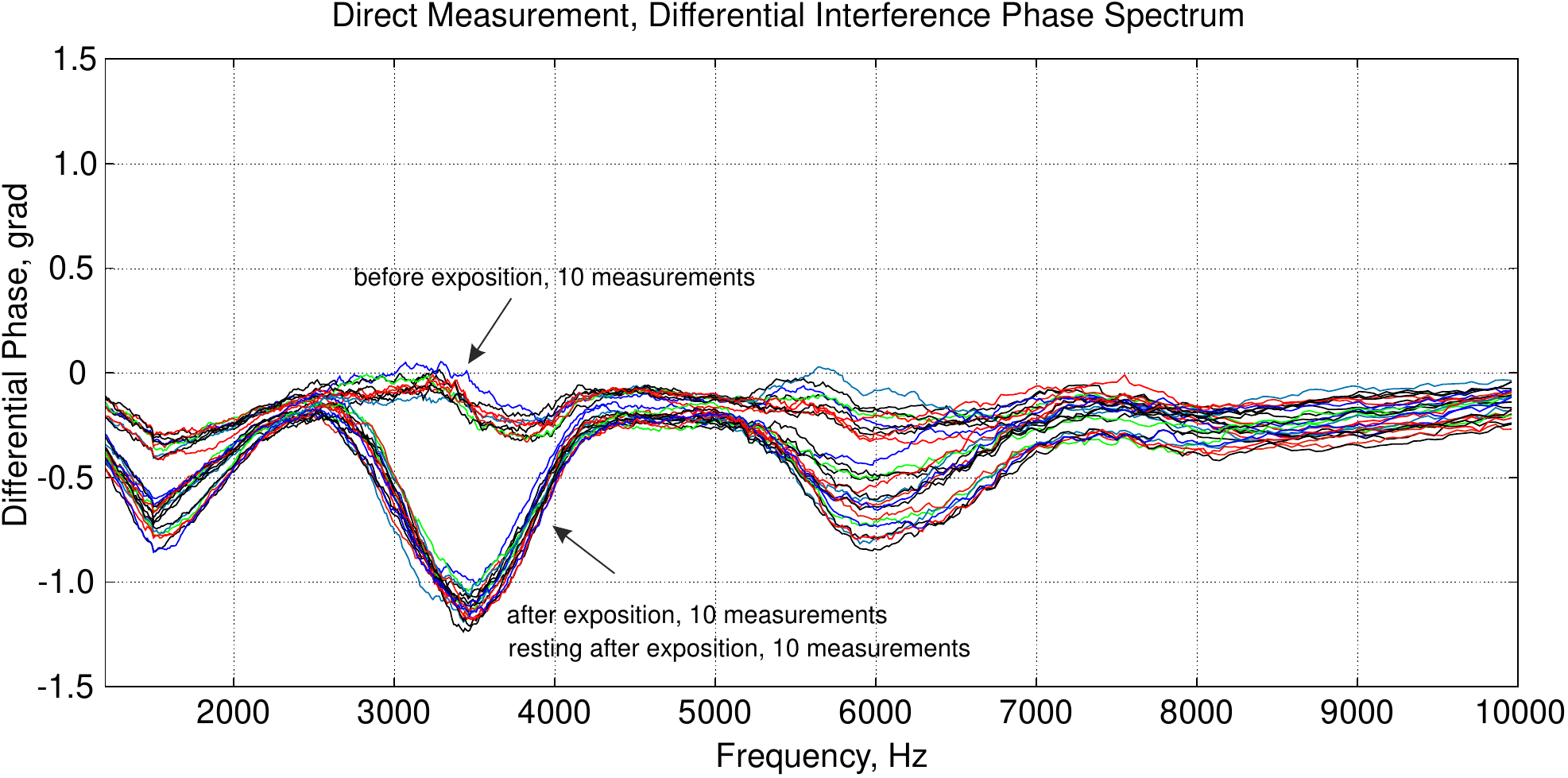}}
\subfigure[]{\includegraphics[width=.49\textwidth]{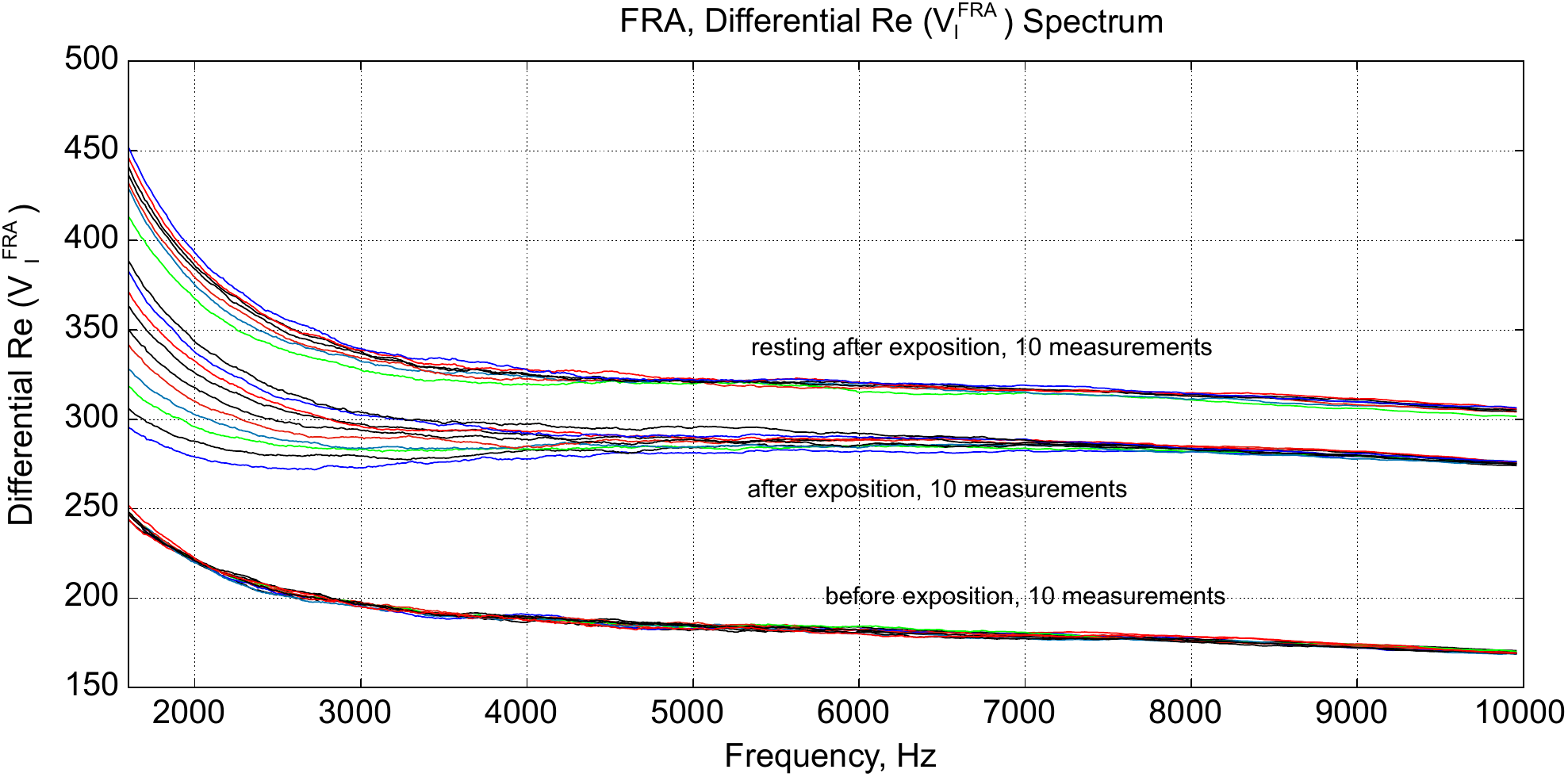}}~
\subfigure[]{\includegraphics[width=.49\textwidth]{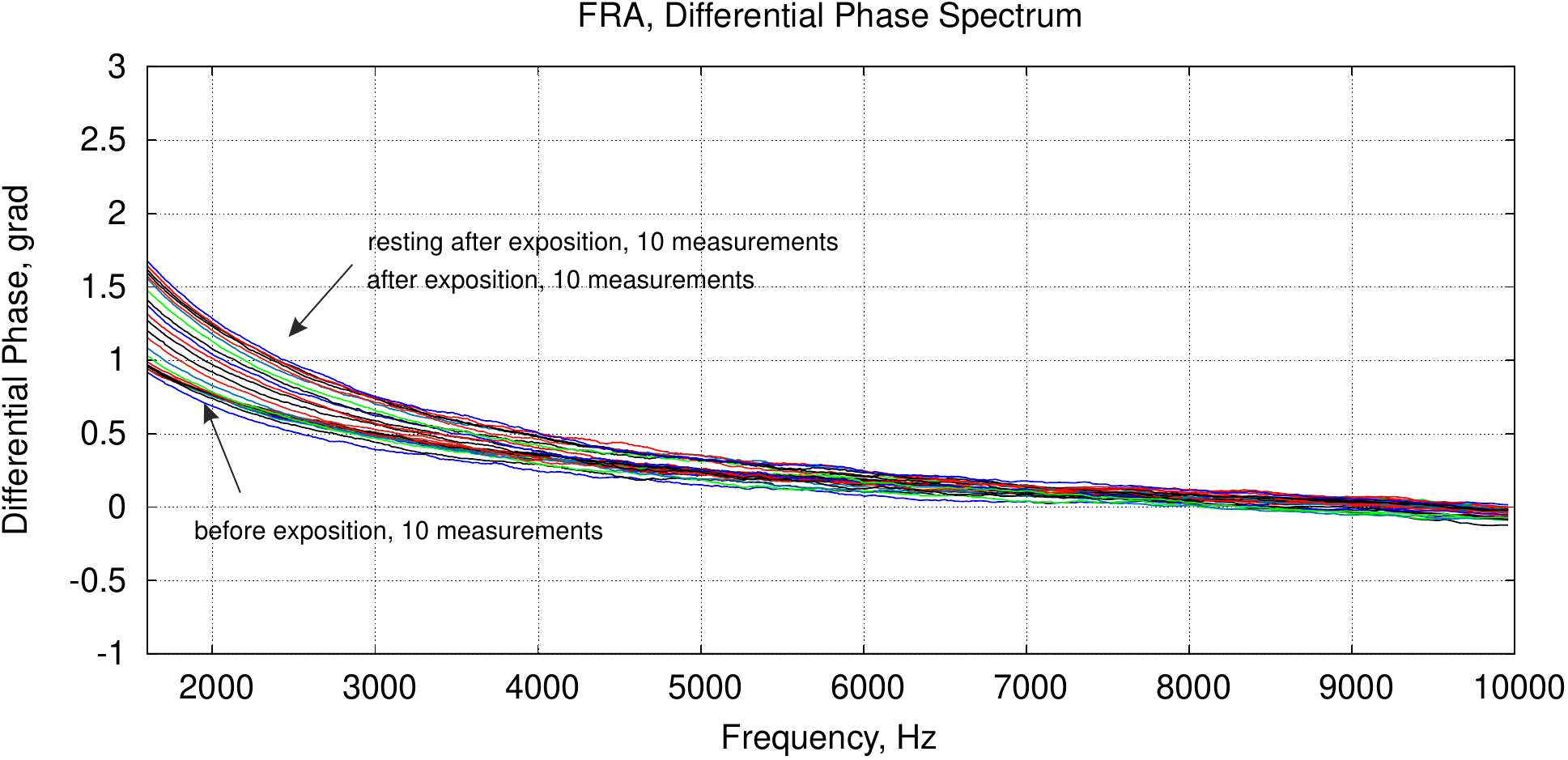}}
\caption{\small The experimental results with water exposed in the \emph{'Cosma'} generator, excitation time before measurement is 1 ms, the MU-EIS system is used. \textbf{(a,b)} Direct measurements of amplitude and phase characteristics, \textbf{(c,d)} Results of frequency response analysis.
\label{fig:measurementWater2}}
\end{figure*}

\begin{figure*}[hbp]
\centering
\subfigure[]{\includegraphics[width=.49\textwidth]{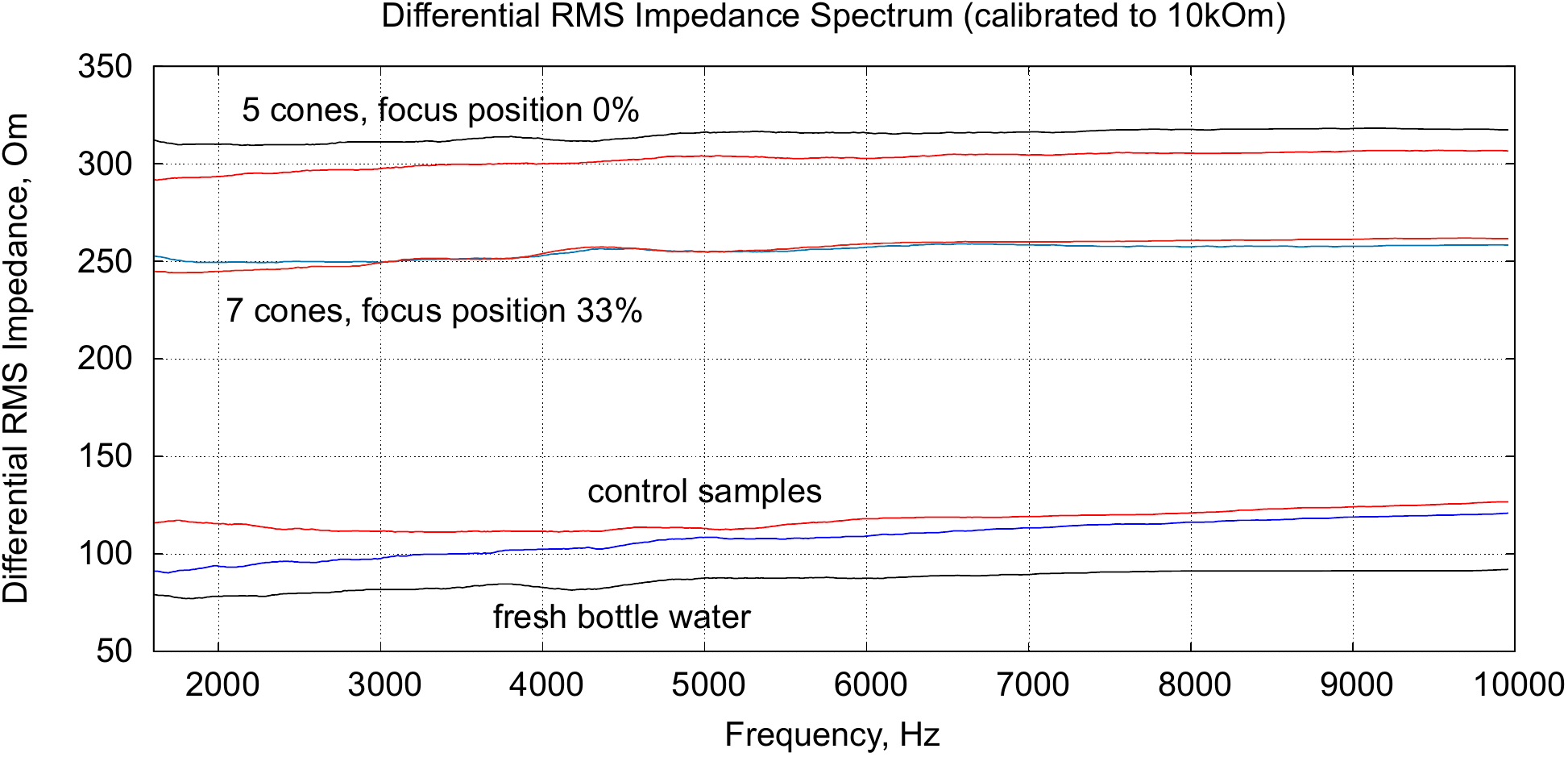}}~
\subfigure[]{\includegraphics[width=.49\textwidth]{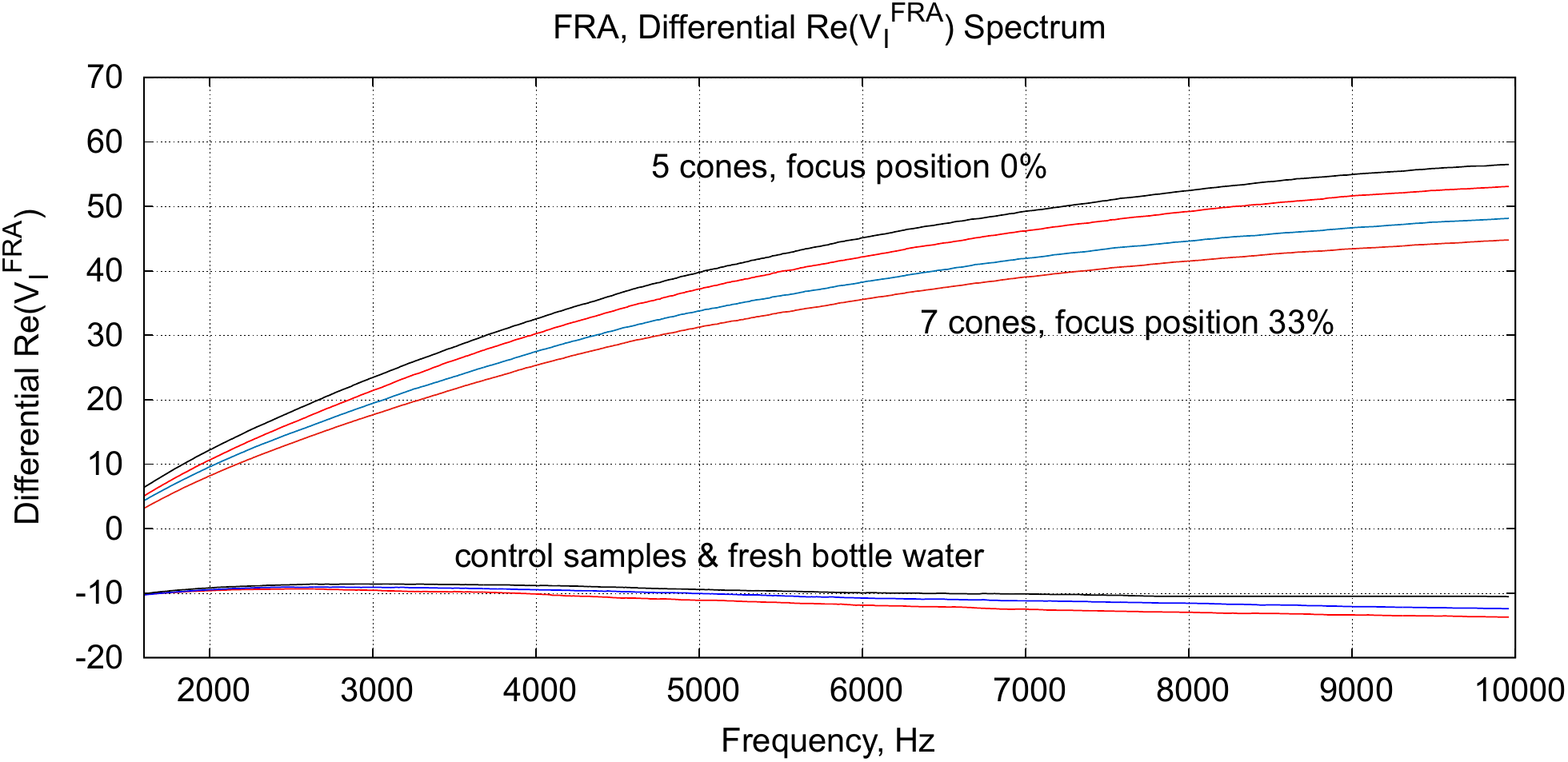}}
\subfigure[]{\includegraphics[width=.49\textwidth]{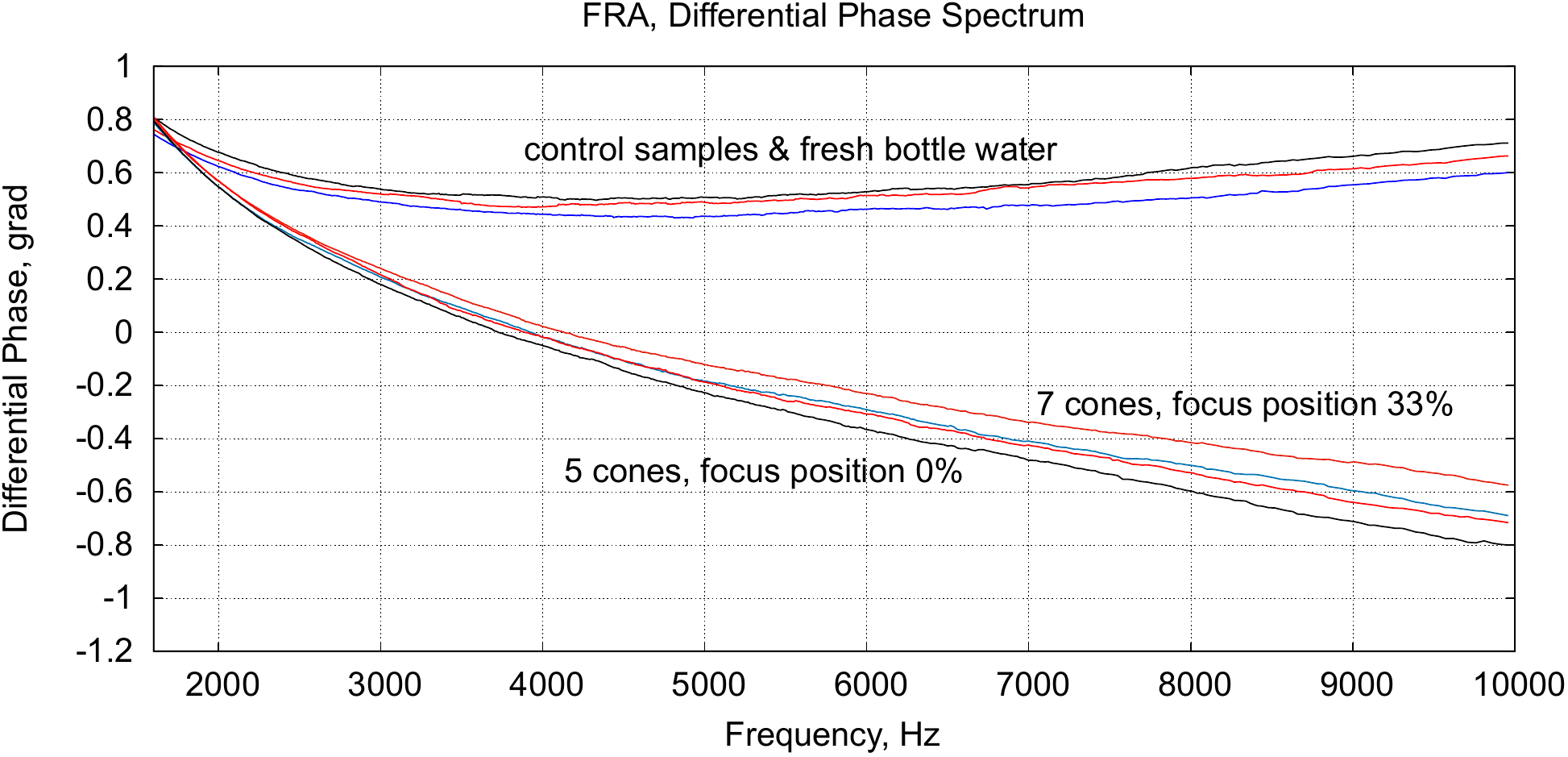}}~
\subfigure[]{\includegraphics[width=.49\textwidth]{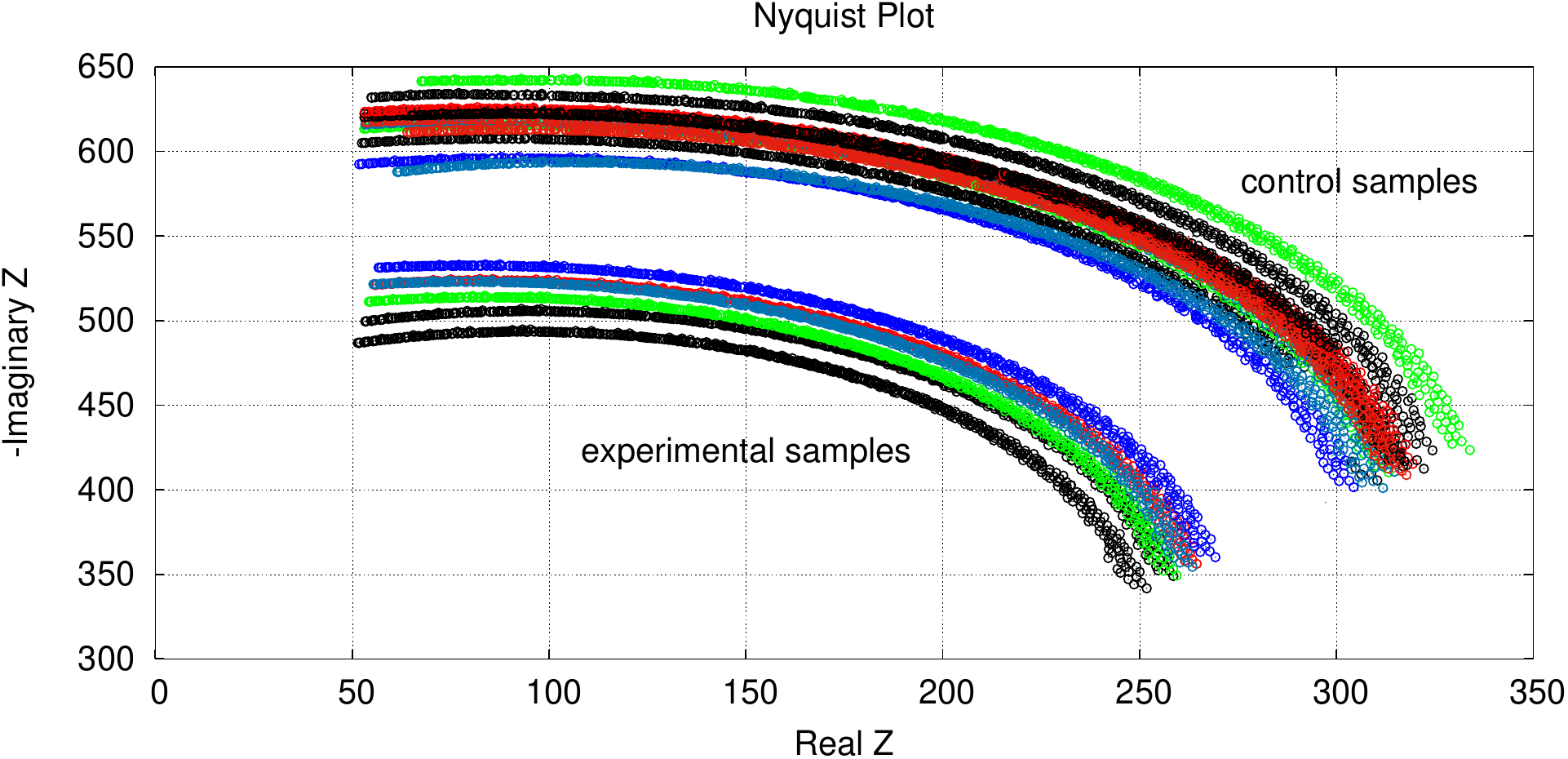}}
\caption{\small The results of EIS analysis of water samples placed in front of passive cone-shaped geometric structures with focus position of 0\% and 33\%. \textbf{(a)} Differential RMS impedance (calibrated to 10k), \textbf{(b)} FRA, the differential spectrum of $Re(V_I)$ impedance, \textbf{(c)} FRA, differential phase spectrum, \textbf{(d)} the Nyquist plot. Thermostat temperature was set to 28$^\circ C \pm 0.02^\circ C$.
\label{fig:konus}}
\end{figure*}

\begin{figure*}[ht]
\centering
\subfigure[]{\includegraphics[width=.49\textwidth]{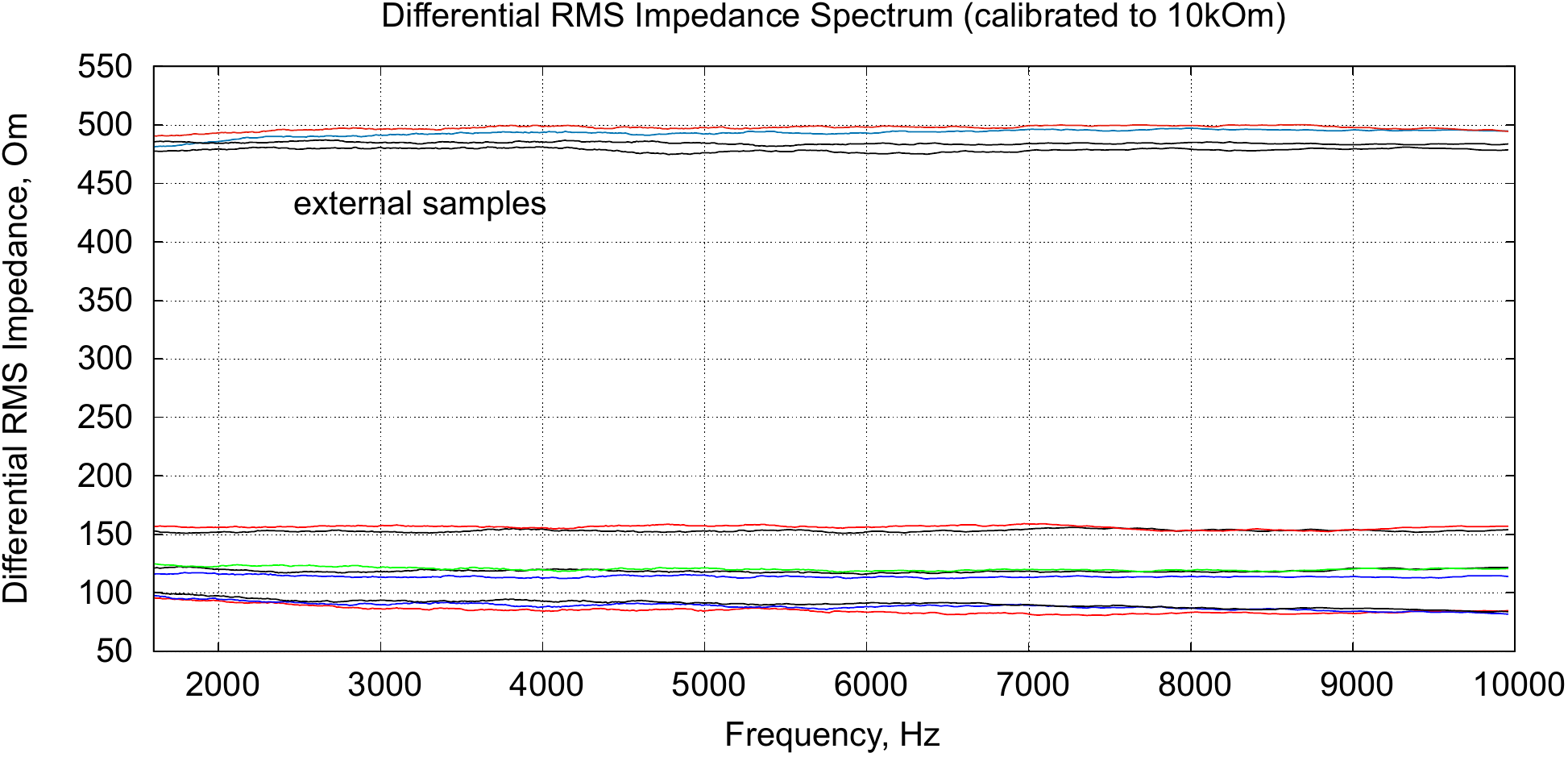}}~
\subfigure[]{\includegraphics[width=.49\textwidth]{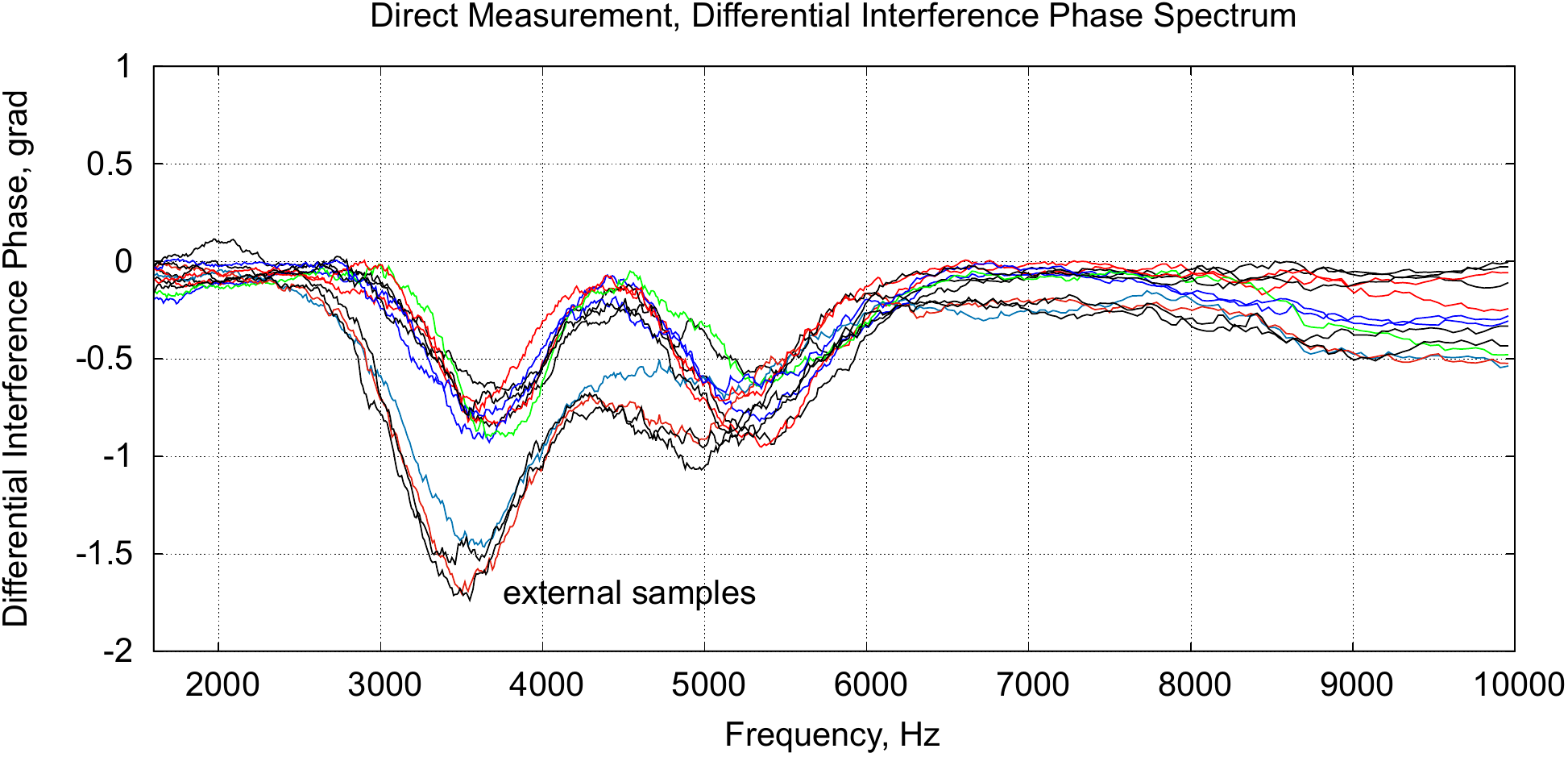}}
\subfigure[]{\includegraphics[width=.49\textwidth]{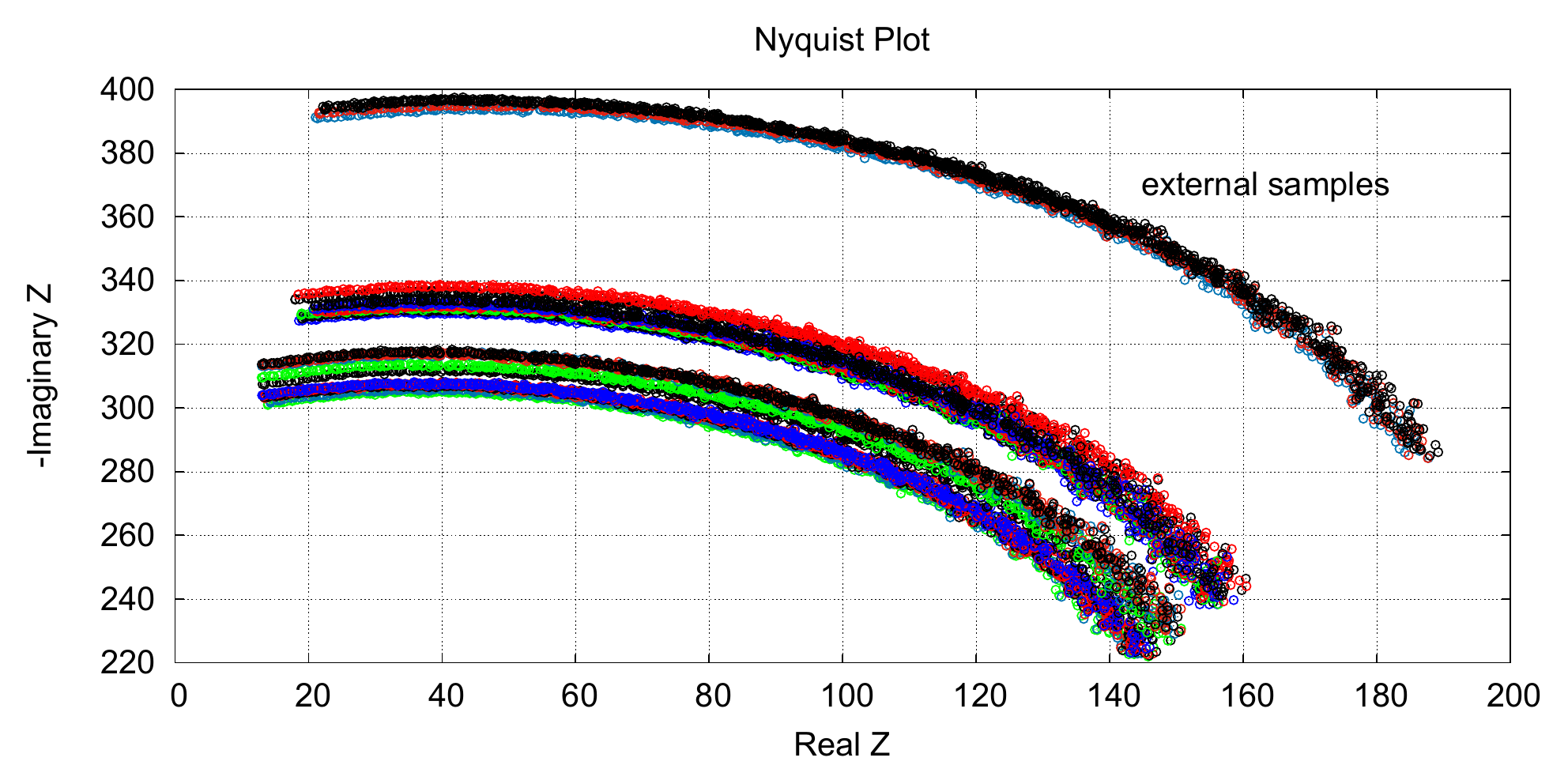}}~
\subfigure[]{\includegraphics[width=.49\textwidth]{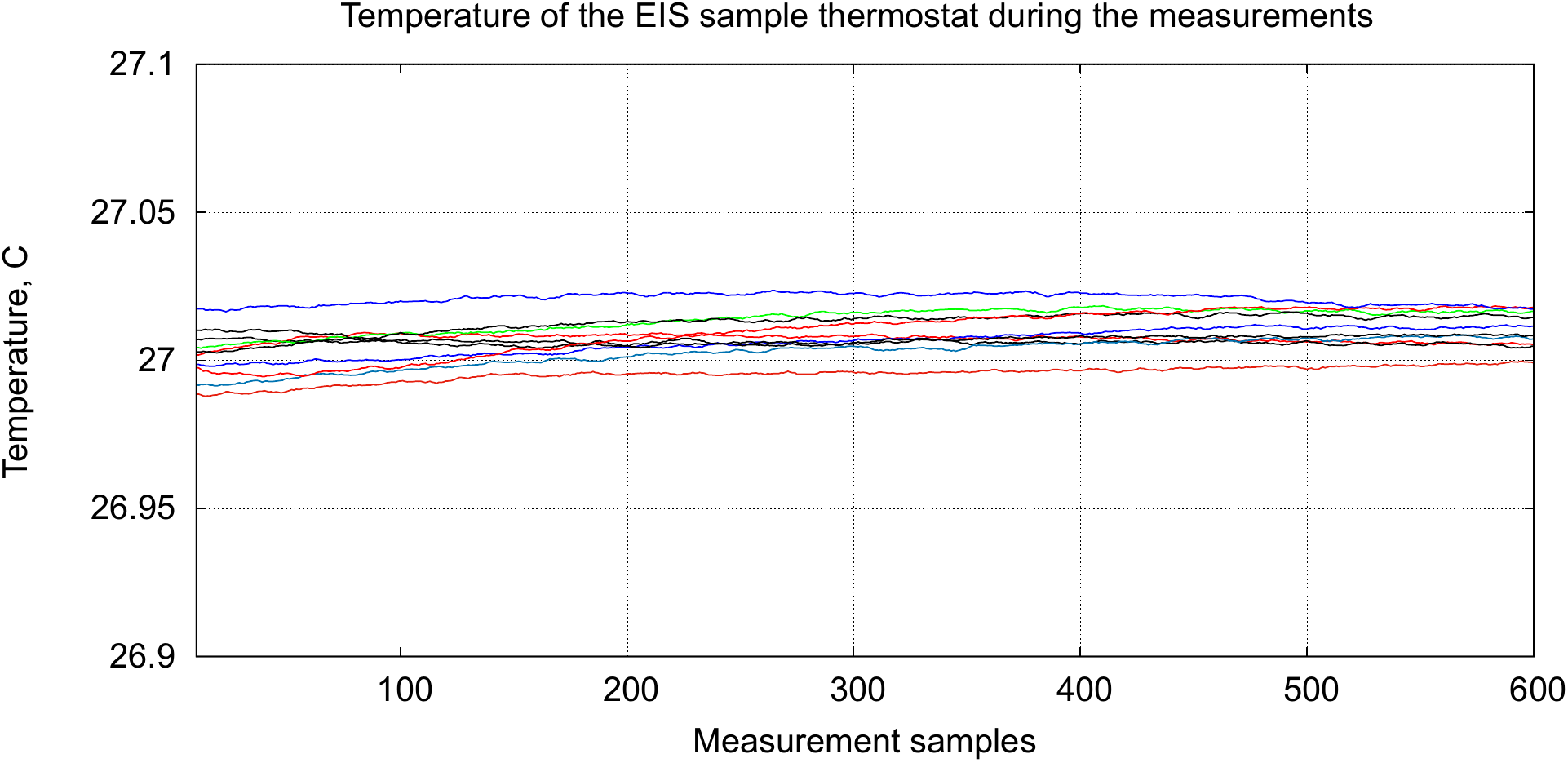}}
\caption{\small The results of EIS analysis of 4 water samples placed for 12 hours in different locations, the 'external sample' -- the water sample placed in the far room (2.5 km from the lab) that shows a strong deviation of the measured EIS parameters, each measurement is repeated twice. \textbf{(a)} Differential range RMS impedance (calibrated to 10k), \textbf{(b)} the differential spectrum of the interference phase shift, \textbf{c)} the Nyquist plot, \textbf{(d)} thermostat temperature with samples during the measurement (the set temperature is 27$^\circ$ C).
\label{fig:measurementGeoBio}}
\end{figure*}

\textbf{Experimental Series 1}. Figure \ref{fig:calibration} shows the results of one of the control experiments with the procedure (1). The amplitude and phase characteristics show small changes caused by variation of initial conditions. These experiments are repeated more than 20 times.

\textbf{Experimental Series 2}. This series has several dozen of iterations, where various parameters of the module \emph{'Cosma'} are tested. The results of one of these experiments is shown in Fig. \ref{fig:measurementWater2}, where the procedures (2) and (3) are applied -- water samples are exposed and the random error is estimated after the exposure. The exposition time varies between 10 and 30 minutes. The variation of random error before and after the exposure has a similar character. However, significant changes are observed in the exposed samples. The measurement results can be characterized by $\Delta V_I^{diff}$ -- amplitude changes of the differential signal ($Re(V_I^{FRA})$, $Re(Z)$, $M(f)$, $Corr(f)$ exhibit similar changes), $\Delta \Phi$ - change of the interference phase shift and $\Delta V_I^{stationary}$ -- variation of stationary conditions. Repeated measurements of the irradiated samples after 6-24 hours show a strong variation at low frequencies that can indicate a change of electrochemical stability.

\textbf{Experimental Series 3}. These experiments are conducted with the cone-shaped geometric structures. Containers with water are placed in front of the output cone for 36 hours, see Fig. \ref{fig:coneSample}. Control measurements are performed with fresh bottle water as well as with control containers rested for 36 hours without any impact. Results of several measurements are shown in Fig. \ref{fig:konus}. There are almost no differences between control samples and fresh bottle water, however, an essential difference between experimental and control samples. Since all containers are positioned in one room with the same EM and other environmental factors, we can assume a non-electromagnetic impacting factor related to the shape effect.

\textbf{Experimental Series 4}. For this series, four identical samples of water in 15 ml containers are placed for 12 hours in different locations with the distance between 3 meters up to 2.5 km from each other. This experiment is repeated 4 times. Fig. \ref{fig:measurementGeoBio} shows the EIS results for one of the experiments. In addition, it shows the temperature of samples during measurements, the temperature fluctuations do not exceed $0.02^\circ$C. The 'external sample' (sample from the far location) demonstrates the largest changes of EIS parameters. Since the measured values of EM and radiation background in these locations are on the same level and conform to the EC/DIN norms, we can explain the EIS differences of samples only by some geological or geo-biological factor.

\begin{figure}[ht]
\centering
\subfigure{\includegraphics[width=.4\textwidth]{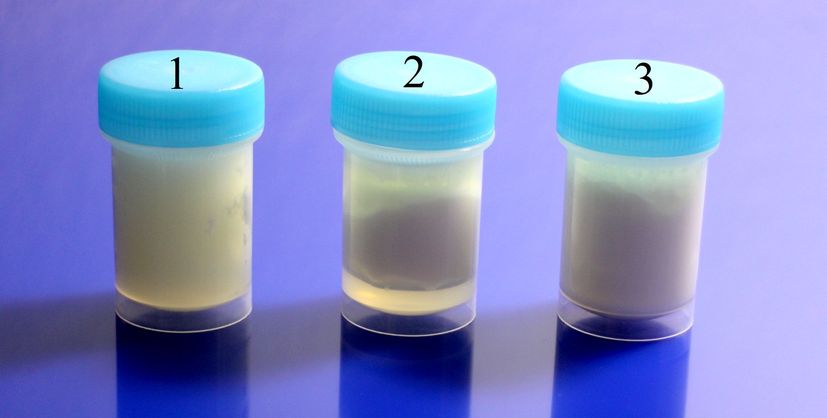}}~
\caption{\small The milk samples two weeks after irradiation, the container 1 has an experimental sample,  the containers 2 and 3 -- control samples. Clotted milk in the container 1 differs significantly from 2 and 3.
\label{fig:milk2weeks}}
\end{figure}

\textbf{Experimental Series 5.} This series of experiments is motivated by the works of V.A.Sokolova's group \cite{Sokolova2002en}, where a gel-like consistency of milk exposed by the A.Deev' generator was achieved. Moreover, the well-known patents of R.Pavlita \cite{PavlitaFoundation92}, A.Deev's work \cite{Rubel13en} and ISTC VENT (group of A.E.Akimov) \cite{Kernbach13arXiv} indicated a capability of cleaning suspended solutions exposed to some sources of weak emissions. For example, Fig. \ref{fig:milk2weeks} shows samples of milk two weeks after irradiation in the module \emph{'Cosma'}, we indeed observe a difference between experimental and control samples of the clotted milk.

\begin{figure*}[ht]
\centering
\subfigure[]{\includegraphics[width=.49\textwidth]{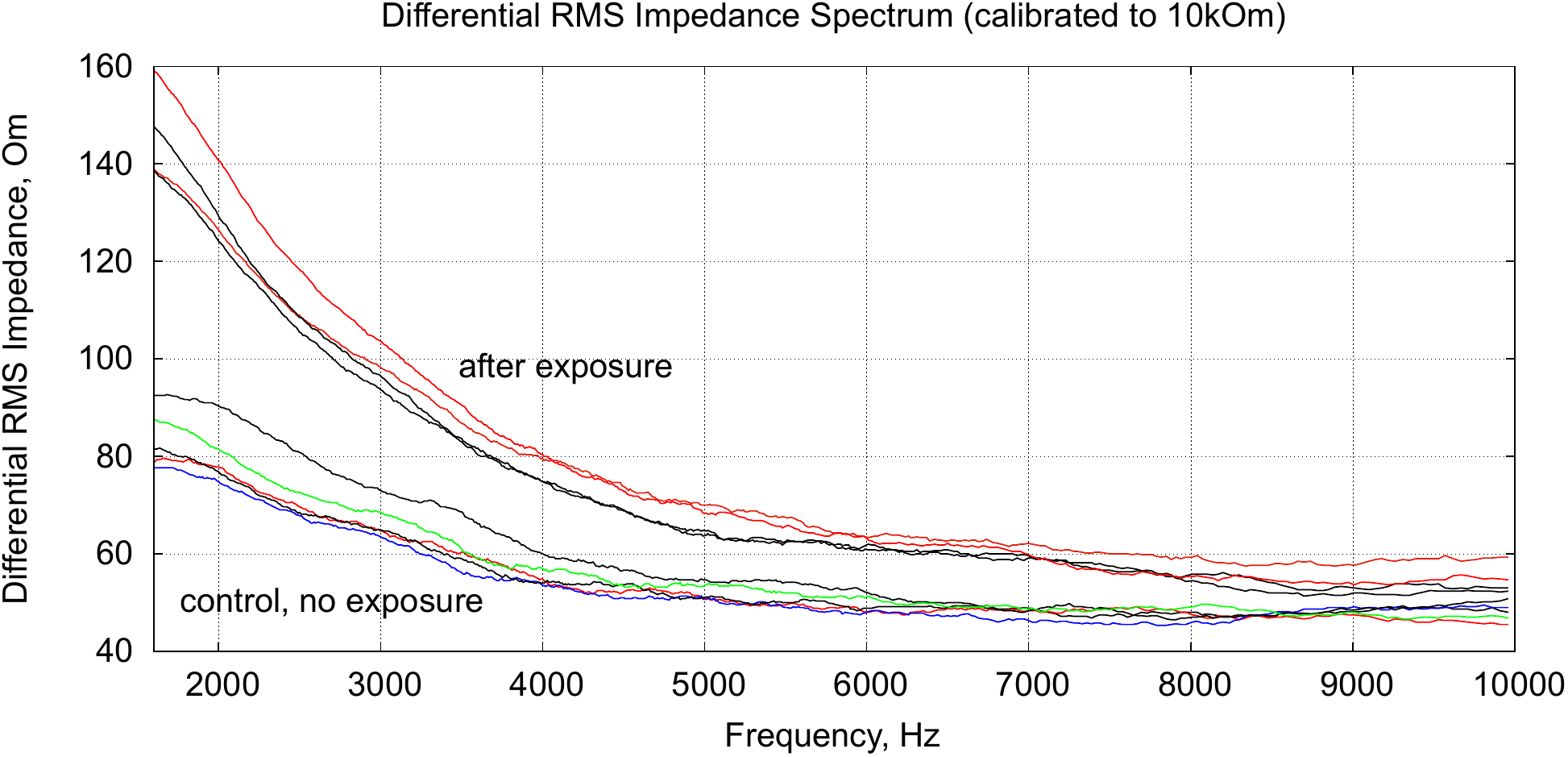}}~
\subfigure[]{\includegraphics[width=.49\textwidth]{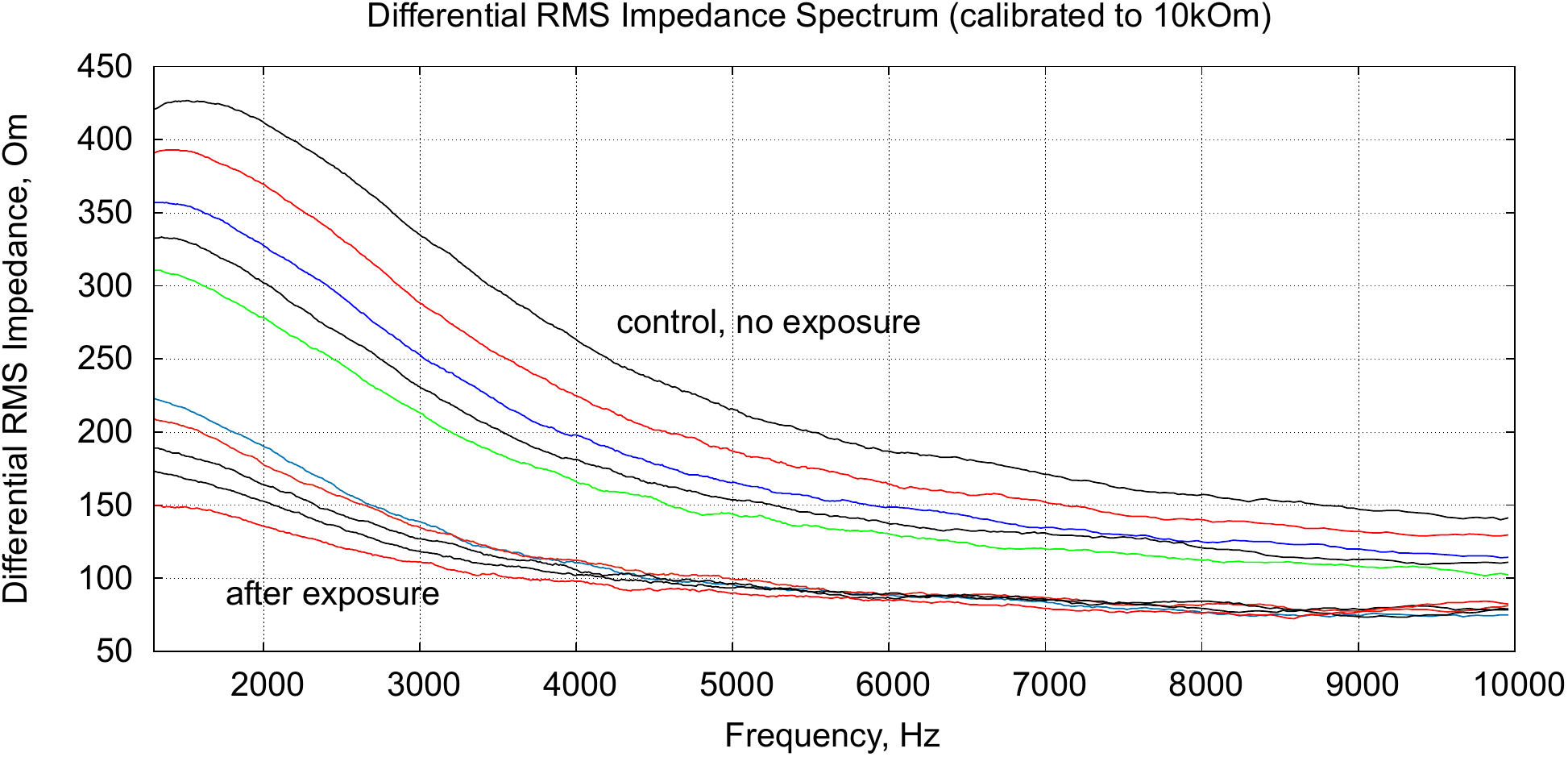}}
\subfigure[]{\includegraphics[width=.49\textwidth]{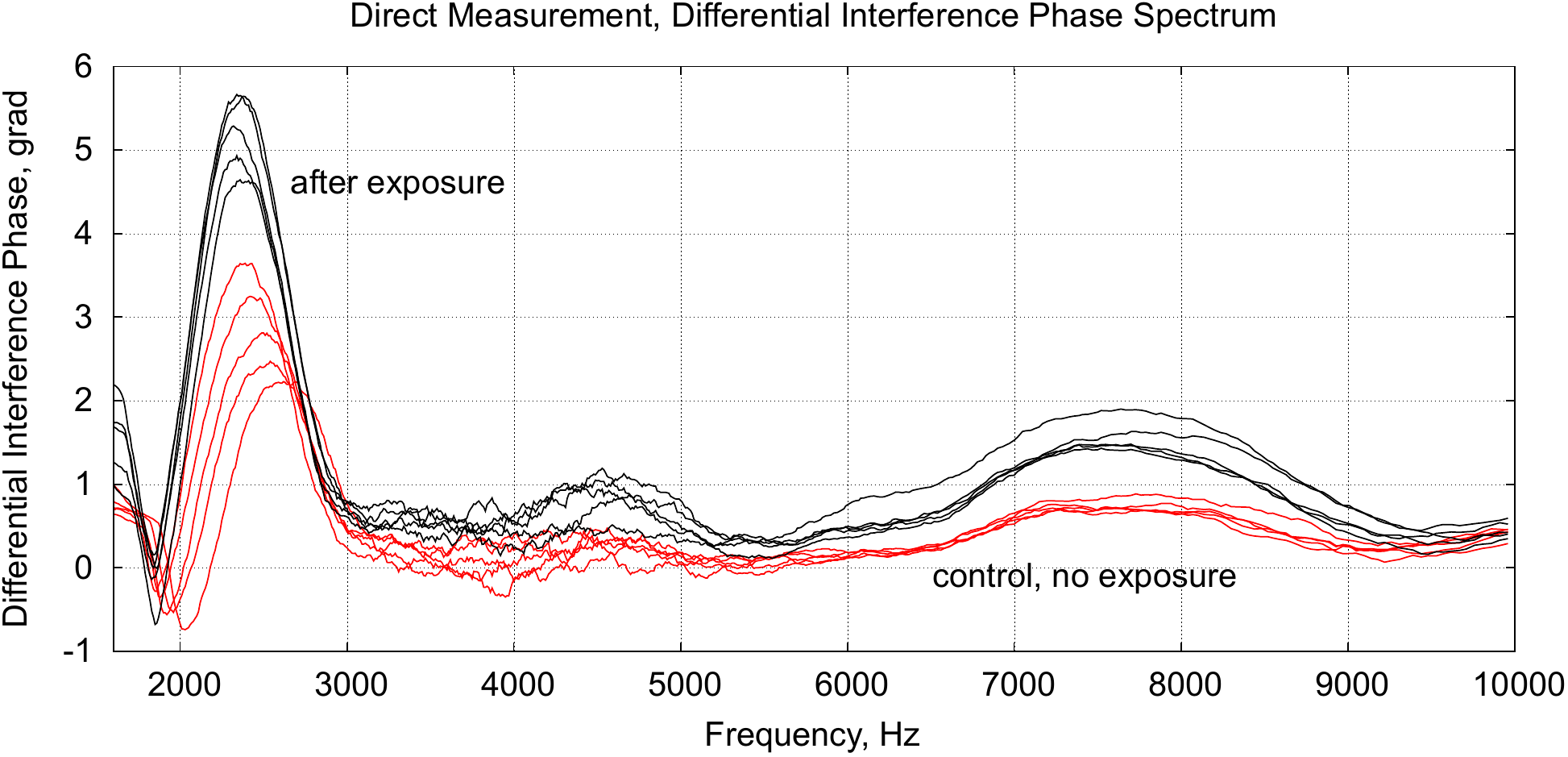}}~
\subfigure[]{\includegraphics[width=.49\textwidth]{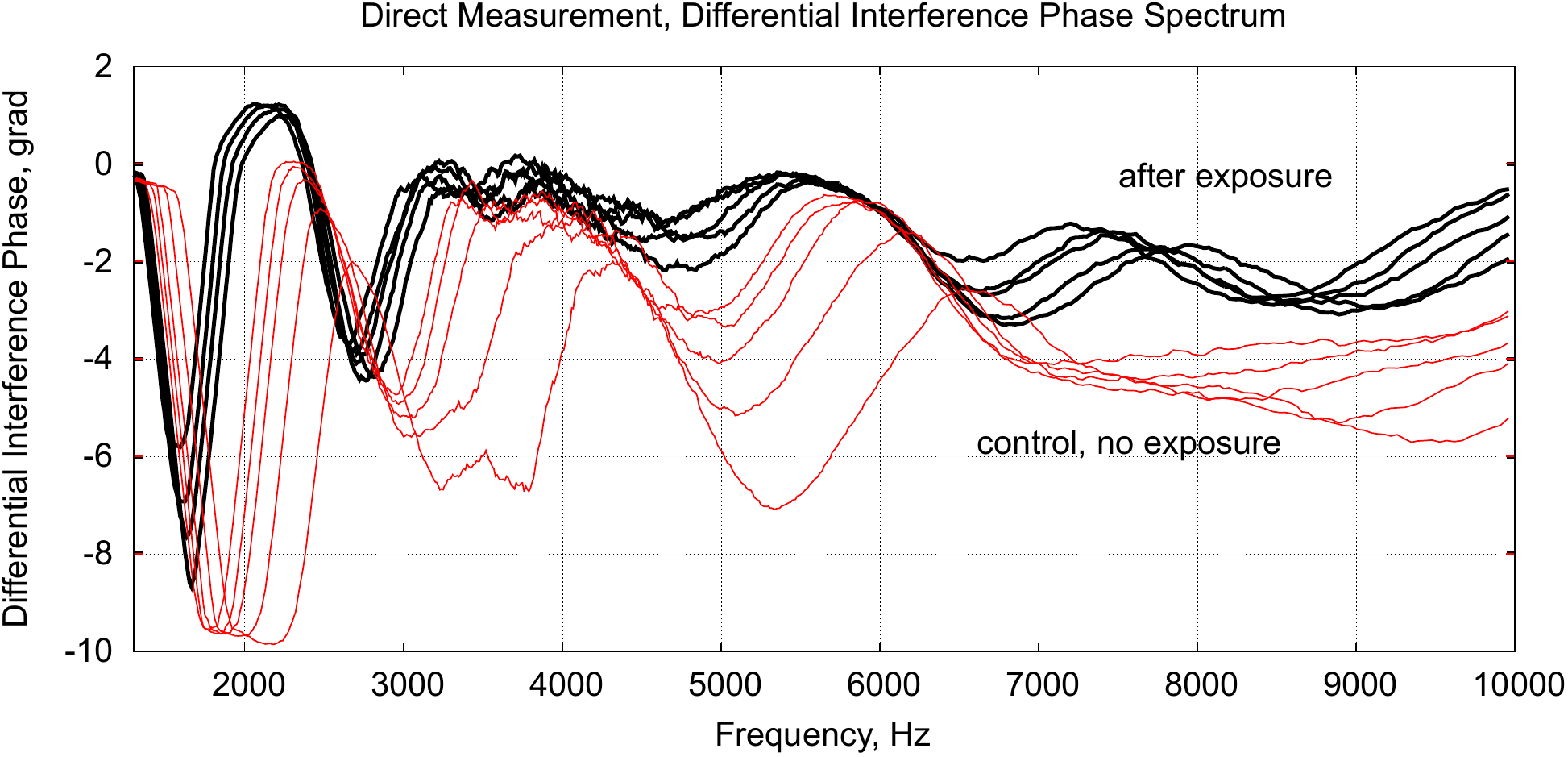}}
\caption{\small The results of EIS analysis of milk samples, 5 replicate measurements are shown in each case, \textbf {(a, c)} -- the samples immediately after irradiation, \textbf{(b, d)} -- the sample 24 hours after irradiation, the samples are stored at room temperature. \textbf{(a, b)} Differential impedance spectrum RMS (calibrated to 10k); \textbf{(c, d)} spectrum of differential interference phase shift. The thermostat temperature is set to $27^\circ C \pm 0.02^\circ C$, as shown in Fig. \ref{fig:measurementGeoBio}(d).
\label{fig:measurementMilk}}
\end{figure*}

Experiments are performed with 1.5\% milk from different manufacturers and are repeated 4 times. The results are shown in Fig.  \ref{fig:measurementMilk}. Similarly to water samples, we observe a change of impedance and interference phase shift of irradiated samples. However, in these experiments, a strong variation of electrochemical stability before, after exposure and 24 hours after exposure is observed. Apparently, the complex organic compounds have their own dynamics, caused by biochemical processes. For example, determining the freshness of milk by measuring its conductivity is well known. In this sense, milk is not a 'good EIS marker' for analyzing weak emissions.

\section{Conclusion}
\label{sec:conclusion}

This paper demonstrates an accurate differential approach with electrochemical impedance spectroscopy adapted for analysis of weak emissions. This method provides reliable results with a high degree of repeatability. In experiments with samples exposed to artificial and natural EM/non-EM emission, the measurement results allow distinguishing treated and untreated samples in all cases. Due to a short measurement time, this method is potentially suitable for a rapid analysis in field conditions. Other applications, e.g. in autonomous systems \cite{Levi99,Eiben:2011:EAE:2001858.2001874} and robotics \cite{Kornienko_S05e,Kornienko_OS01} are also possible.

The values of $\Delta V_I^{diff} $ (differential signal amplitude), $\Delta \Phi$ (interference phase shift), $Re(Z), Im(Z)$ (for example, the Nyquist plot) and changes in electrochemical stationary describe differences between control and experimental samples. To some extent, these value can characterize the exposure by weak emission. Based on the Randles's electrochemical model \cite{Randles47}, as shown in Fig. \ref{fig:scheme}, these parameters indicate a change in the near-electrode layer parameters and diffusion processes associated with Warburg impedance. 

Three types of measurements are performed: test measurements with the variation of initial conditions, impact by experimental devices, and analysis of samples taken from different geological locations. In all cases, the differential measurements (comparison with control untreated samples) are used and the parameters $\Delta V_I^{diff}$, $\Delta \Phi$ and $Re(Z), Im(Z)$ characterized the impact. Results of impact are more 'evident', if the measurements are performed 12-24 hour after exposition.

The greatest measurement error is caused by the variation of initial conditions and electrochemical stability. Since the EIS is an 'invasive' method of analysis, i.e. this method interacts with samples during measurement, not all liquids and not all voltages $V_V$ are suitable for the analysis of weak interactions. It is necessary to find a fluid with a high electrochemical stability and a good response to weak emissions. This liquid will act as an 'EIS marker'.

Experiments with commercial devices based on AD5933 showed three major disadvantages: lack of temperature stabilization, the inability of differential measurements and FRA analysis with window functions. The resulting error of such measurements are often higher than the amplitude of measured signals caused by the impact of experimental factors. In particular, oscillations caused by the Hanning function in AD5933 complicate the differential analysis. Similarly to the potentiometry \cite{Kernbach14dpHen}, \cite{Kernbach15dpHen}, is necessary to develop instruments specifically adapted for such measurements.

In preparing and conducting the experiments, similar works of other authors are analyzed. In particular publications of pioneers of Soviet unconventional studies V.A.Sokolova et al \cite{Sokolova2002en} is  considered. The well-known publications of A.E.Akimov' group \cite{Akimov01en} are also referring to these works. We can confirm that the largest changes in impacted samples relate to amplitude parameters that are measured by Sokolova as a relative dispersion of conductivity. In general terms, we think that the replication of those experiments is successful, however, several issues remained open. For example, the used fluid and organic tissues posses different electrical conductivity. It is not clear how the impedance matching was performed. Sokolova also obtained different conductivity values at frequencies from 1 kHz to 8 kHz for the same material. In our experiments, however, the differences in conductivity are much lower. This might indicate considerable measurement errors in Sokolova' experiments, e.g. caused by manual insertion of electrodes and missing temperature stabilization.

In further studies we will collect larger statistics for various liquid, suspensions and gels, as well as for methods and devices generating weak emissions.

\small

\begin{thebibliography}{10}

\bibitem{Chang10}
Byoung-Yong Chang and Su-Moon Park.
\newblock Electrochemical impedance spectroscopy.
\newblock {\em Annu. Rev. Anal. Chem.}, 3:207--29, 2010.

\bibitem{Ganesh08}
V.~Ganesh, R.R. Pandey, B.D. Malhotra, and V.~Lakshminarayanan.
\newblock Electrochemical characterization of self-assembled monolayers (sams)
  of thiophenol and aminothiophenols on polycrystalline au: effects of
  potential cycling and mixed sam formation.
\newblock {\em J. Electroanal. Chem}, 619:87--97, 2008.

\bibitem{Macdonald06}
D.D. Macdonald.
\newblock Reflections on the history of electrochemical impedance spectroscopy.
\newblock {\em Electrochim. Acta}, 51:1376--88, 2006.

\bibitem{Puthoff98}
H.E. Puthoff.
\newblock Communication method and apparatus with signals comprising scalar and
  vector potentials without electromagnetic fields.
\newblock {\em Patent US5845220}, 1998.

\bibitem{AkimovPatent92en}
A.E. Akimov, V.Ch. Tarasenko, A.V. Samochin, I.V. Kurick, V.P. Meiboroda, V.A.
  Licharev, and U.F. Perov.
\newblock {\em Patent SU1748662 Approach and device for correction of
  structural properties of materials (rus)}.
\newblock 1992.

\bibitem{Burgin08}
Luc B\"urgin.
\newblock {\em Der Urzeit-Code}.
\newblock Herbig, 2008.

\bibitem{Kernbach12JSE}
Serge Kernbach.
\newblock Replication attempt: Measuring water conductivity with polarized
  electrodes.
\newblock {\em Journal of Scientific Exploration}, 27(1):69--105, 2013.

\bibitem{Kumar05}
I.R.Kumar, N.V.C.Swamy, and H.R.Nagendra.
\newblock Effect of pyramids on microorganisms.
\newblock {\em Indian Journal of Traditional Knowledge}, 4(4):373--379, 2005.

\bibitem{Makin02en}
S.V. Mjkin, I.V. Vasilieva, and A.V. Rudenko.
\newblock Investigation of the influence of the field generated by a pyramid on
  the material objects (rus).
\newblock {\em {C}onsciousness and physical reality}, (7(2)):45--53, 2002.

\bibitem{Dunne95}
Brenda~J. Dunne and Robert~G. Jahn.
\newblock Consciousness and anomalous physical phenomena.
\newblock {\em Technical Note PEAR 95004}, 1995.

\bibitem{Schmidt71}
H.~Schmidt.
\newblock Mental influence on random events.
\newblock {\em New Scientist and Science Journal}, pages 757--758, 1971.

\bibitem{Tompkins73}
Peter Tompkins and Christopher Bird.
\newblock {\em The Secret Life of Plants}.
\newblock Hardcover, 1973.

\bibitem{Bobrov97en}
V.~Bobrov.
\newblock {R}eaction of double electrical layer on torsion field (rus).
\newblock In {\em BINITI N 1055-B97}, 1997.

\bibitem{Bobrov06en}
A.V. Bobrov.
\newblock {\em {I}nvestigating a field concept of consciousness (rus)}.
\newblock Orel, Orel University Publishing, 2006.

\bibitem{Cardella01}
C.~Cardella, L.~de~Magistris, E.~Florio, and C.W. Smith.
\newblock Permanent changes in the physico-chemical properties of water
  following exposure to resonant circuits.
\newblock {\em Journal of Scientific Exploration}, (15(4)):501--518, 2001.

\bibitem{Stenschke1985261}
H.~Stenschke.
\newblock Polarization of water in the metal/electrolyte interface.
\newblock {\em Journal of Electroanalytical Chemistry and Interfacial
  Electrochemistry}, 196(2):261 -- 274, 1985.

\bibitem{F29837900225}
David W.~R. Gruen and Stjepan Marcelja.
\newblock Spatially varying polarization in water. a model for the electric
  double layer and the hydration force.
\newblock {\em J. Chem. Soc.{,} Faraday Trans. 2}, 79:225--242, 1983.

\bibitem{doi:10.1021/la00077a011}
M.~L. Belaya, M.~V. Feigel'man, and V.~G. Levadnyii.
\newblock Structural forces as a result of nonlocal water polarizability.
\newblock {\em Langmuir}, 3(5):648--654, 1987.

\bibitem{Lyklema05}
J.~Lyklema.
\newblock {\em Fundamentals of Interface and Colloid Science}.
\newblock Academic Press, 2005.

\bibitem{Sokolova2002en}
V.A.Sokolova.
\newblock {\em First experimental confirmation of torsion fields and their
  usage in agriculture (rus)}.
\newblock Moscow, 2002.

\bibitem{Andriasheva15en}
M.A. Andriasheva.
\newblock Changing water properties through numeric codes (rus).
\newblock {\em IJUS}, 10(3):7--14, 2015.

\bibitem{krasn10en}
V.G.Krasnobrygev and M.V.Kurick.
\newblock Properties of coherent water (rus).
\newblock {\em Quantum Magic}, 7(2):2161--2166, 2010.

\bibitem{Krinker122en}
Mark Krinker.
\newblock Spinning process based info-sensors.
\newblock {\em Proc. of the 3rd int. conf. 'Torsion fields and information
  interactions'}, pages 223--228, 2012.

\bibitem{6223212}
M.~Krinker, A.~Goykadosh, and H.~Einhorn.
\newblock On the possibility of transferring information with
  non-electromagnetic fields, the relation of spinning processes and encoding
  information and the hydrogen spin detector.
\newblock In {\em Systems, Applications and Technology Conference (LISAT), 2012
  IEEE Long Island}, pages 1--12, May 2012.

\bibitem{Kernbach14minimalen}
S.~Kernbach.
\newblock The minimal experiment (rus).
\newblock {\em International Journal of Uncoventional Science}, 4(2):50--61,
  2014.

\bibitem{Kernbach14dpHen}
S.Kernbach and O.Kernbach.
\newblock On precise \emph{pH} and \emph{dpH} measurements (rus).
\newblock {\em International Journal of Unconventional Science}, 5(2):83--103,
  2014.

\bibitem{Kernbach15dpHen}
S.~Kernbach and O.~Kernbach.
\newblock Detection of ultraweak interactions by precision dph approach (rus).
\newblock {\em IJUS}, 9(3):17--41, 2015.

\bibitem{Anosov03en}
V.N. Anosov and E.M. Truchan.
\newblock New approach to impact of weak magnetic fields on living objects
  (rus).
\newblock {\em {Doklady Akedemii Nauk}: biochemistry, biophysics and molecular
  biology}, (392):1--5, 2003.

\bibitem{Randles47}
J.E.B. Randles.
\newblock Kinetics of rapid electrode reactions.
\newblock {\em Discuss. Faraday Soc.}, 1:11--19, 1947.

\bibitem{Tkachuk10en}
U.V. Tkachuk, S.D.Jremchuk, and A.A.Fedotov.
\newblock Experimental study of the effect of rotating ferrite magnetic disks
  on the reaction of acetic anhydride hydration (rus).
\newblock {\em The II int. conf. 'Torsion fields and information
  interactions'}, pages 106--110, 2010.

\bibitem{ImpedanceAgilent}
Agilent Technologies.
\newblock {\em Agilent Impedance Measurement Handbook}.
\newblock Agilent, 2013.

\bibitem{Norouzi11}
P.Norouzi, M.Pirali-Hamedani, T.M.Garakani, and M.R.Ganjali.
\newblock Application of fast fourier transforms in some advanced
  electroanalytical methods.
\newblock {\em in: Fourier Transforms -- New Analytical Approaches and FTIR
  Strategies, (ed.) G.Nikolic}, pages 303--322, 2011.

\bibitem{Chabowski15}
K.~Chabowski, T.Piasecki, A.Dzierka, and K.~Nitsch.
\newblock Simple wide frequency range impedance meter based on {AD5933}
  integrated circuit.
\newblock {\em Metrology and Measurement Systems}, XXII(1):13--24, 2015.

\bibitem{1375091}
Jong-Wook Kim, ByungKoo Park, Seung~Cheol Jeong, Sang~Woo Kim, and PooGyeon
  Park.
\newblock Fault diagnosis of a power transformer using an improved
  frequency-response analysis.
\newblock {\em Power Delivery, IEEE Transactions on}, 20(1):169--178, Jan 2005.

\bibitem{Matsiev15}
L.Matsiev.
\newblock Improving performance and versatility of systems based on
  single-frequency dft detectors such as ad5933.
\newblock {\em Electronics}, 4(1):1--34, 2015.

\bibitem{Ojarand13}
J.~Ojarand and M.~Min.
\newblock Simple and efficient excitation signals for fast impedance
  spectroscopy.
\newblock {\em Elektronika ir Elektrotechnika}, 19(2):1392--1215, 2013.

\bibitem{Mejna09}
A.Mejía-Aguilar and R.Pallàs-Areny.
\newblock Electrical impedance measurement using voltage/current pulse
  excitation.
\newblock {\em XIX IMEKO World Congress, Fundamental and Applied Metrology},
  pages 662--667, 2009.

\bibitem{Smith76}
D.E. Smith.
\newblock The acquisition of electrochemical response spectra by on-line fast
  fourier transform.
\newblock {\em Data processing in electrochemistry. Anal. Chem.}, 48:A221--40,
  1976.

\bibitem{Kernbach13metrologyen}
S.~Kernbach.
\newblock On metrology of systems operating with 'high-penetrating' emission
  (rus).
\newblock {\em International Journal of Unconventional Science}, 1(2):76--91,
  2013.

\bibitem{Harris78}
F.J. Harris.
\newblock On the use of windows for harmonic analysis with the discrete fourier
  transform.
\newblock {\em IEEE Proc.}, 66:51–83, 1978.

\bibitem{AD5933}
Analog Devices.
\newblock {\em Data sheet AD5933: 1 MSPS, 12-Bit Impedance Converter, Network
  Analyzer}.
\newblock Analog Devices, 2005-2013.

\bibitem{0957-0233-24-10-102001}
Paul~Ben Ishai, Mark~S Talary, Andreas Caduff, Evgeniya Levy, and Yuri Feldman.
\newblock Electrode polarization in dielectric measurements: a review.
\newblock {\em Measurement Science and Technology}, 24(10):102001, 2013.

\bibitem{Kalvoy11}
Kalvoy H, Johnsen GK, Martinsen OG, and Grimnes S.
\newblock New method for separation of electrode polarization impedance from
  measured tissue impedance.
\newblock {\em The Open Biomedical Engineering Journal}, 5:8--13, 2011.

\bibitem{Kernbach12ITen}
S.~Kernbach.
\newblock Exploration of high-penetrating capability of {LED} and laser
  emission. {Parts} 1 and 2 (rus).
\newblock {\em Nano- and microsystem's technics}, 6,7:38--46,28--38, 2013.

\bibitem{Kernbach13formsen}
Serge Kernbach and Olga Kernbach.
\newblock Impact of structural elements on high frequency non-contact
  conductometry.
\newblock {\em in publication}, 2013.

\bibitem{Ghaffari15}
S.A. Ghaffari, W.-O. Caron, M.~Loubier, M.~Rioux, J.~Viens, B.~Gosselin, and
  Y.~Messaddeq.
\newblock A wireless multi-sensor dielectric impedance spectroscopy platform.
\newblock {\em Sensors}, 15:23572--23588, 2015.

\bibitem{Hoja2010191}
Jerzy Hoja and Grzegorz Lentka.
\newblock Interface circuit for impedance sensors using two specialized
  single-chip microsystems.
\newblock {\em Sensors and Actuators A: Physical}, 163(1):191 -- 197, 2010.

\bibitem{Veinik81en}
A.I.Weinik.
\newblock {\em Book of sorrow (rus)}.
\newblock Minsk manuscript, 1981.

\bibitem{Veinik91en}
A.I.Weinik.
\newblock {\em Thermodynamics of real processes (rus)}.
\newblock Minks: 'Nauka i technika', 1991.

\bibitem{Chigevsky73en}
A.L.Chigewskij.
\newblock {\em Electric and magnetic properties of erythrocytes (rus)}.
\newblock Moscow, 1973.

\bibitem{Chigevsky30en}
A.L.Chigewskij.
\newblock {\em Epidemiological catastrophes and periodic activity of the Sun}.
\newblock Moscow, 1930.

\bibitem{Dulnev04en}
G.Dulnev.
\newblock {\em Looking for a new world (rus)}.
\newblock Wes, 2004.

\bibitem{Asheulov00en}
A.A. Asheulow, U.B. Dobrovolskij, and V.A. Besulik.
\newblock Impact of electric and magnetic fields on parameters of semiconductor
  devices.
\newblock {\em Technology and design in electronic equipment}, (1):33--35,
  2000.

\bibitem{brit98en}
A.A.Britova, I.V.Adamko, and V.L.Bachurina.
\newblock Activation of water by laser light, magnetic field or by their
  combination (rus).
\newblock {\em Vesnik of Novgorod's State University}, (7), 1998.

\bibitem{Kernbach15en}
S.~Kernbach.
\newblock {\em Supernatural. Scientifically proven facts}.
\newblock Algorithm. Moscow, 2015.

\bibitem{PavlitaFoundation92}
The~Pavlita Foundation.
\newblock {\em Note on work of Robert Pavlita and his experiments in
  bio-energy}.
\newblock www.keelynet.com/biology/pavlita1.txt, 1992.

\bibitem{Rubel13en}
Edwin~C. May, Victor Rubel, and Loyd Auerbach.
\newblock {\em ESP WARS: East and West: An Account of the Military Use of
  Psychic Espionage As Narrated by the Key Russian and American Players}.
\newblock CreateSpace Independent Publishing Platform, 2014.

\bibitem{Kernbach13arXiv}
S.~{Kernbach}.
\newblock {Unconventional research in USSR and Russia: short overview}.
\newblock {\em arXiv 1312.1148}, 2013.

\bibitem{Levi99}
P.~Levi, M.~Schanz, S.~Kornienko, and O.~Kornienko.
\newblock Application of order parameter equation for the analysis and the
  control of nonlinear time discrete dynamical systems.
\newblock {\em Int. J. Bifurcation and Chaos}, 9(8):1619--1634, 1999.

\bibitem{Eiben:2011:EAE:2001858.2001874}
A.~E. Eiben, N.~Ferreira, M.~C. Schut, and S.~Kernbach.
\newblock Embodied artificial evolution: the future of artificial evolutionary
  systems.
\newblock In {\em Proc. of the 13th conf. on Genetic and evolutionary
  computation}, GECCO '11, pages 27--28, New York, NY, USA, 2011. ACM.

\bibitem{Kornienko_S05e}
S.~Kornienko, O.~Kornienko, and P.~Levi.
\newblock Swarm embodiment - a new way for deriving emergent behaviour in
  artificial swarms.
\newblock In P.Levi and et~al., editors, {\em Autonome Mobile Systeme
  (AMS'05)}, pages 25--32, 2005.

\bibitem{Kornienko_OS01}
O.~Kornienko, S.~Kornienko, and P.~Levi.
\newblock Collective decision making using natural self-organization in
  distributed systems.
\newblock In {\em Proc. of Int. Conf. on Computational Intelligence for
  Modelling, Control and Automation (CIMCA'2001), Las Vegas, USA}, pages
  460--471, 2001.

\bibitem{Akimov01en}
A.E. Akimov, V.J. Tarasenko, and S.U. Tolmachev.
\newblock Torsion communication -- new system for telecommunication (rus).
\newblock {\em Electrocommunication}, (5), 2001.

\end{thebibliography}

\end{document}